\documentclass{cernrep}
\pdfoutput=1 

\RequirePackage[T1]{fontenc}
\RequirePackage[utf8]{inputenc}
\RequirePackage{times}
\RequirePackage{xspace}

\RequirePackage[usenames,dvipsnames,svgnames,x11names,table]{xcolor}
\RequirePackage{lineno}  
\RequirePackage{tabularx}
\RequirePackage{graphicx}  
\RequirePackage{amsmath} 
\RequirePackage{amssymb}
\RequirePackage{amsfonts}
\RequirePackage{upgreek} 

\RequirePackage[font=footnotesize,labelfont=bf]{caption}
\RequirePackage[font=footnotesize,labelfont=bf]{subcaption}

\RequirePackage{doi}
\RequirePackage{hyperref}

\newcommand{\mathDP}{\gamma^{\mathrm{D}}}
\newcommand{\mathmDP}{m_{\mathDP}}
\newcommand{\DP}{$\mathDP$\xspace}
\newcommand{\mDP}{$\mathmDP$\xspace}
\newcommand{\Br}{\mathop\mathrm{Br}}

\begin{document}

\title{Sensitivity of the SHiP experiment to dark photons decaying to a pair of charged particles}

\author{SHiP Collaboration\thanks 
       {e-mails: a.magnan@imperial.ac.uk, atakan.tugberk.akmete@cern.ch (corresponding authors)}}

\institute{}

\begin{abstract}
  Dark photons are hypothetical massive vector particles that could
  mix with ordinary photons. The simplest theoretical model is fully
  characterised by only two parameters: the mass of the dark photon
  m$_{\gamma^{\mathrm{D}}}$ and its mixing parameter with the photon,
  $\varepsilon$. The sensitivity of the SHiP detector is reviewed for
  dark photons in the mass range between 0.002 and 10\,GeV. Different
  production mechanisms are simulated, with the dark photons decaying
  to pairs of visible fermions, including both leptons and
  quarks. Exclusion contours are presented and compared with those of
  past experiments.  The SHiP detector is expected to have a unique
  sensitivity for m$_{\gamma^{\mathrm{D}}}$ ranging between 0.8 and
  3.3$^{+0.2}_{-0.5}$\,GeV, and $\varepsilon^2$ ranging between
  $10^{-11}$ and $10^{-17}$.
\end{abstract}

\keywords{High Energy Physics; fixed target experiment; CERN; intensity frontier; hidden sector; dark photons; beam dump facility}

\maketitle

\section{Introduction}
\label{sec:intro}

The CERN beam facility located near Geneva, Switzerland, comprises
several particle accelerators among which the Super Proton Synchrotron
(SPS) and the Large Hadron Collider (LHC)~\cite{LHC}. The SPS is an
essential part of the accelerator chain delivering 400\,GeV proton
beams to the LHC but also to fixed-target experiments. The LHC is
planned to be upgraded into a high-luminosity machine starting
operation around 2026 with the HL-LHC
program~\cite{Apollinari:2015bam}. In parallel to the high-energy
frontier probed by the LHC, a complementary way of exploring the
parameter space of potential new physics is through the ``intensity
frontier''. The SPS physics programme is hence proposed to be further
extended via the construction of a beam dump facility
(BDF)~\cite{Ahdida:2655435}. The BDF foresees the full exploitation of
the SPS accelerator, which with its present performance could allow
the delivery of up to $4 \times 10^{19}$ protons on target per year,
while respecting the beam requirements of the HL-LHC and maintaining
the operation of the existing SPS beam facilities.

By probing lower-energy scenarios with high-intensity beams, the aim
is to identify whether new physics could be hidden from sight due to
weak connections through portals instead of direct interactions with
the known particles, with the new particles belonging to a hidden
sector. The simplest renormalisable extensions of the standard model
(SM) are possible through three types of
portals~\cite{PhysRevD.80.095024,Alekhin_2016}, involving either a
scalar (e.g. dark Higgs boson~\cite{Patt:2006fw,PhysRevD.75.037701}),
a vector (e.g. dark photon~\cite{Holdom:1985ag,Bauer:2018onh}) or
fermions (e.g. heavy neutral leptons~\cite{Gorbunov_2007}). The LHC
experiments have already derived strong constraints on short-lived
high-mass
mediators~\cite{Aaboud:2018fvk,Aaboud:2017buh,Sirunyan:2018mgs,Khachatryan:2016zqb}.
Scenarios with long-lived mediators with relatively low masses however
remain largely unexplored. The SHiP (Search for Hidden Particles)
experiment~\cite{Anelli:2007512} has been proposed in
2013~\cite{Bonivento:1606085} and is designed to look for particles
which would decay in the range 50--120\,m from their production
vertices. The sensitivity of the SHiP detector to heavy neutral
leptons has been investigated in Ref.~\cite{SHiP:2018xqw}. This
article is dedicated to studying the sensitivity of the SHiP detector
to dark photons.

After describing briefly the SHiP detector and its simulation in
Section~\ref{sec:shipdetector}, the model considered for the dark
photon production and decay is reviewed in
Section~\ref{sec:DPprod}. The sensitivity of the SHiP detector in the
minimal dark photon model with decays to charged particles is given in
Section~\ref{sec:sensitivity} for the three production modes
studied. Finally Section~\ref{sec:concl} provides a conclusion.

\section{The SHiP detector and simulation}
\label{sec:shipdetector}

SHiP~\cite{Anelli:2007512} is a new general purpose fixed-target
experiment intended to exploit the proposed BDF to search for
particles present in hidden portal models. The 400\,GeV proton beam
extracted from the SPS will be dumped on a high density target with
the aim of accumulating $2\times 10^{20}$ protons on target during 5
years of operation. A dedicated detector, based on a long vacuum tank
followed by a spectrometer and particle identification detectors, will
allow probing a variety of models with light long-lived exotic
particles and masses below $\mathcal O(10)\,\rm{GeV}$. A critical
component of SHiP is the muon shield, which deflects the high flux of
muons produced in the target~\cite{SHiP:2020hyy,Akmete:2017bpl}, that
would represent a serious background in the search for hidden-sector
particles. The detector is designed to fully reconstruct the exclusive
decays of hidden particles and to reject the background down to below
0.1 events in the sample of $2\times 10^{20}$ protons on
target~\cite{Ahdida:2704147}.

The detector consists of a large magnetic spectrometer located
downstream of a 50 m-long and up to $5\times 11$ m-wide cone-shaped
decay volume~\cite{Miano2020}. To suppress the background from
neutrinos interacting in the fiducial volume, the decay volume is
maintained under a vacuum.  The spectrometer tracker is designed to
accurately reconstruct the decay vertex, mass and impact parameter of
the decaying particle. A set of calorimeters followed by muon chambers
provide identification of electrons, photons, muons and charged
hadrons. A dedicated timing detector measures the coincidence of the
decay products, which allows the rejection of combinatorial
backgrounds. The decay volume is surrounded by background taggers to
tag neutrino and muon inelastic scattering in the surrounding
structures, which may produce long-lived SM $\rm{V}^0$ particles, such
as $\rm{K}_{\rm{L}}$, that have topologies similar to the expected
signals.

\begin{figure}[h!]
  \begin{minipage}[b]{0.68\textwidth}
    \includegraphics[width=1.\linewidth]{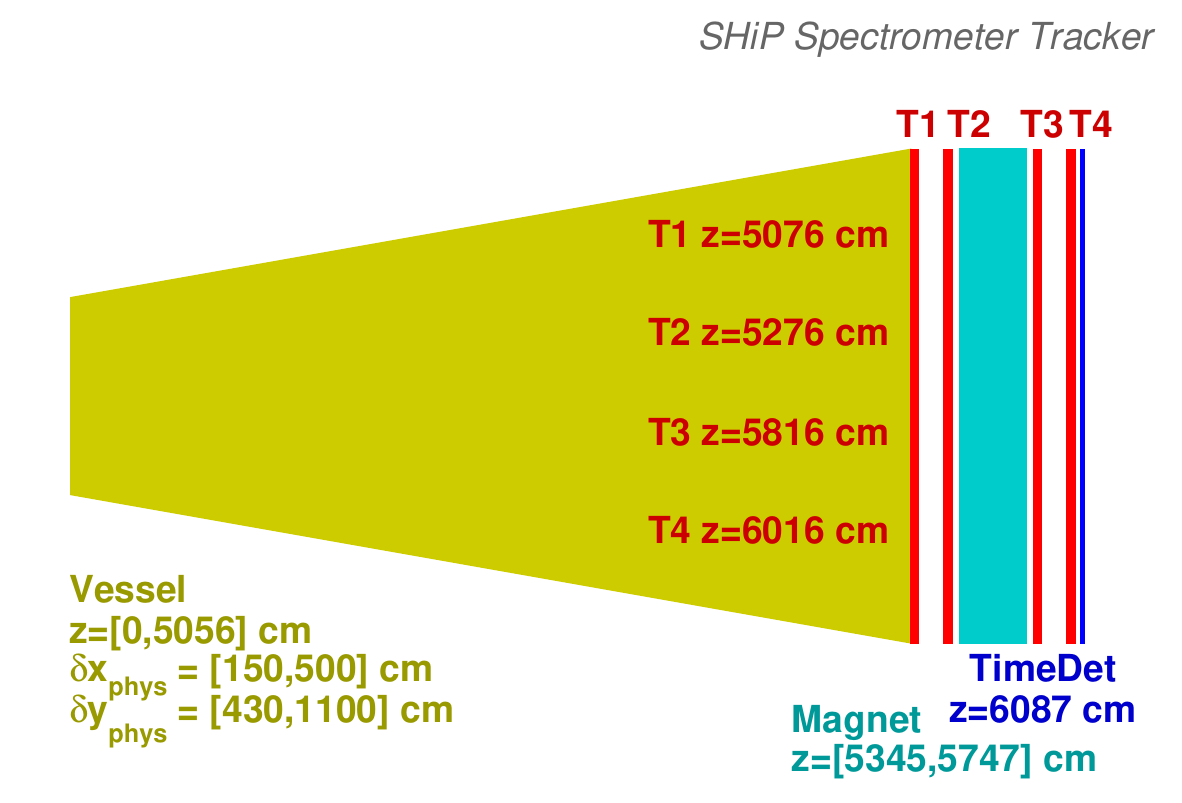}
    \caption{Positions of the vessel and tracking stations along the beam axis (z). The vessel dimensions $\delta x_{\mathrm{phys}}$ and $\delta y_{\mathrm{phys}}$ represent the upstream and downstream physics acceptance in the plane transverse to the beam axis.}
    \label{fig:stations} 
  \end{minipage}
  \hfill
  \begin{minipage}[b]{0.27\textwidth}
    \includegraphics[width=1.\linewidth]{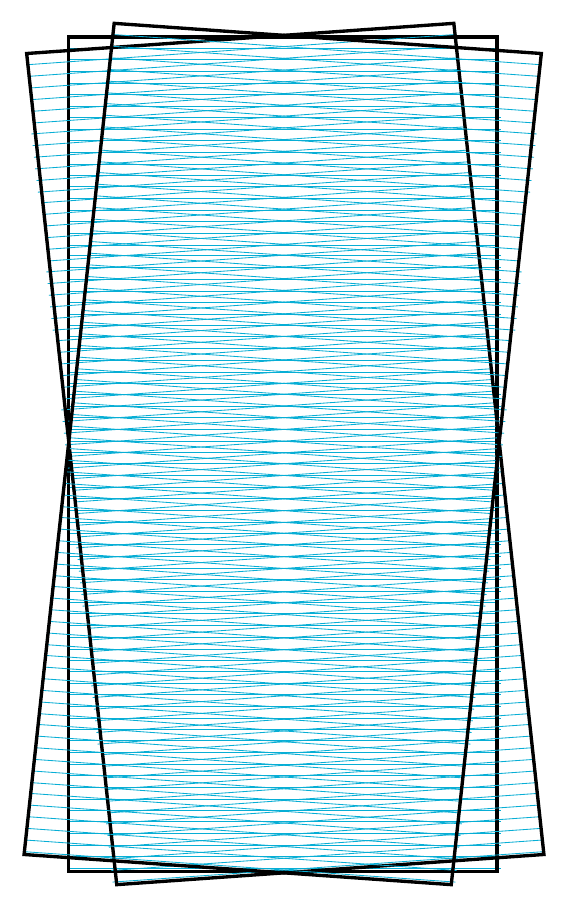}
    \caption{Schematic drawing of the three ``views'' that compose each straw chamber.}
    \label{fig:chambers}
  \end{minipage}
\end{figure}

The spectrometer tracker is a crucial component in the reconstruction
of the charged particles produced by the decay of dark photons. The
baseline layout consists of four tracking stations (T1 to T4)
symmetrically arranged around a dipole magnet as shown in
Fig.~\ref{fig:stations}. The transverse size of the tracker stations
matches the size of the magnet. Each station consists of 9072 straw
tubes which are arranged in four views (Y-U-V-Y), as shown in
Fig.~\ref{fig:chambers}. The Y view has straws horizontally
aligned. The U and V views are rotated by an angle of
$\theta_{\mathrm{stereo}} = \pm 5^{\mathrm{o}}$. The x coordinate is
hence measured with an accuracy of~1/sin($\theta_{\mathrm{stereo}}$),
directly impacting the measurement of the decay vertex, of the opening
angle of the daughter particles (which enters the invariant mass), and
of the impact parameter at the production target. In order to provide
good spatial resolution and minimise the contribution from multiple
scattering, the straw tubes are made of thin polyethylene
terephthalate (PET). More detail about the initial design of the straw
detector can be found in
Refs.~\cite{Ahdida:2704147,Bereziuk:2005286}. The pattern recognition
algorithms applied to the hits on the straw spectrometer are described
in Ref.~\cite{VanHerwijnen:2005715}, and the algorithms for particle
identification are presented in Ref.~\cite{Hosseini:2282039}.

In the simulation, proton fixed-target collisions are generated by
PYTHIA 8.2~\cite{Sjostrand:2014zea}, inelastic neutrino interactions
by GENIE~\cite{Genie} and inelastic muon interactions by PYTHIA
6~\cite{pythia6}. The heavy-flavour cascade production is also taken
into account~\cite{CERN-SHiP-NOTE-2015-009}. The SHiP detector
response is simulated in the GEANT4~\cite{Geant4} framework. The
simulation is done within FairShip, which is based on the FairRoot
framework~\cite{FAIRROOT}.

\section{Dark photon production and decay}
\label{sec:DPprod}

The minimal dark photon model contains an additional \cal{U}(1) gauge
group to the SM, A$_{\mu}^{\prime}$, whose vector gauge boson is
called the dark photon \DP. A kinetic mixing term between the dark
photon with field strength $F_{\mu\nu}^{\prime}$ and the SM U(1) gauge
bosons with field strength $F^{\mu\nu}_{Y}$ is
allowed~\cite{Holdom:1985ag}, with a reduced strength parameterised by
a coupling $\varepsilon$, also called the kinetic mixing
parameter. The corresponding terms in the Lagrangian can hence be
written as:

\begin{equation}
  \label{eq:Lag}
  \mathcal{L} = \mathcal{L}_{\mathrm{SM}} - \frac{1}{4} (F_{\mu\nu}^{\prime})^2 -\frac{\varepsilon}{2} F_{\mu\nu}^{\prime} F^{\mu\nu}_{Y} + \frac{1}{2} \mathmDP^2 (A_{\mu}^{\prime})^2.
\end{equation}
In its simplest form, the knowledge of the mass of the dark photon
\mDP and the kinetic mixing parameter $\varepsilon$ is enough to
characterise the model and calculate production cross section and
decay properties.

Three different mechanisms are possible for the production of such new
particles at a fixed-target experiment. All of them are studied in
this paper.

The initial 400\,GeV proton beam interacts with the nucleons from the
target material, producing mesons. For meson decay channels involving
photons, the photon can mix with the dark photon, as described in
Section~\ref{sec:meson}. This mode is opened only for dark photon
masses below 0.9\,GeV, as for mesons with masses above this threshold
the main decay channels do not involve photons anymore.

The proton-nucleon interaction could also lead to the radiation of a
dark photon via a bremsstrahlung process, as described in
Section~\ref{sec:pbrem}. This mode is heavily suppressed when the dark
photon mass exceeds that of the proton, and remains relevant only for
masses below $\simeq 2$\,GeV.

The third production mode is via a Drell-Yan like process in Quantum
Chromodynamic (QCD), i.e. quark-antiquark annihilation into the dark
photon, as described in Section~\ref{sec:qcd}. This process is
relevant for dark photon masses in the range
$\mathcal{O}(\hbox{1--10})$\,GeV. Using the parton model with a
factorisation scale below the GeV scale cannot give sensible results,
as expected from the range of validity of parton distribution
functions, and hence this region of the parameter space has not been
considered for this production mechanism.

In this paper, the assumption is made that only the initial proton
interacts. In reality, similar interactions could also happen with
protons or mesons coming from cascade decays happening in the target
material. For electromagnetic processes (electron bremsstrahlung of
photons mixing with the dark photon), it has been shown in
Ref.~\cite{Gorbunov:2014wqa} that their contribution is negligible
compared to the main production mechanisms described above. The study
however remains to be done for hadronic interactions in the cascade
decays, and will be the subject of future work. Hence the expectations
presented here are conservative and the sensitivity could be improved
in the future when this contribution is added.

As a final state, the dark photon decay to pairs of leptons or quarks
as described in Section~\ref{sec:decay} is considered.

\subsection{Production in meson decay}
\label{sec:meson}

The PYTHIA 8.2~\cite{Sjostrand:2014zea} Monte Carlo (MC) generator is
used to produce inclusive QCD events in proton-proton (p-p)
collisions, through all available non-diffractive
processes. Diffractive processes are less important in meson
production, expected to increase the number of mesons produced by
about 15\% according to PYTHIA simulation. Because the diffractive
processes also suffer from larger theoretical uncertainties, they have
not been considered. The leading-order (LO) NNPDF2.3 PDF
set~\cite{Ball:2012cx} has been used with the default Monash 2013
tune~\cite{Skands:2014pea}, and the strong coupling constant set to
$\alpha_{\mathrm{s}} = 0.13$. One proton beam momentum is set to
400\,GeV and the other to 0 (protons or neutrons from the fixed-target
material). The mesons that are produced are then used as sources of
dark photons, if they have decay channels to photons and their decay
to a dark photon of mass \mDP is kinematically allowed. Four processes
are found dominant (with other contributions neglected) and shown in
Table~\ref{tab:mesonDecays}. The decay tables of these four mesons are
reset to having only one decay channel allowed with 100\% branching
ratio ($\pi^{0}\rightarrow \gamma\gamma $, $\eta\rightarrow
\gamma\gamma $, $\omega\rightarrow \pi^{0}\gamma $, $\eta'\rightarrow
\gamma\gamma$). All relevant processes are then added together.

\begin{table}[thbp]
  \caption{Meson decay channels considered for the \DP production. The
    last column shows the average number of mesons expected per
    proton-proton interaction.}
  \label{tab:mesonDecays}
  \centering
  \begin{tabular}{l|c|c|c}
\mDP (GeV) & meson &  $\Br$($\gamma+X$)~\cite{Sjostrand:2014zea} & n$_{\mathrm{meson}}$ / pp \\
\hline 
0--0.135  &    $\pi^{0}\rightarrow\mathDP\gamma$ &  0.98799 &  $6.147 \pm 0.003$ \\
0--0.548  &   $\eta\rightarrow\mathDP\gamma$ &  0.3931181 &  $0.703 \pm 0.008$\\
0--0.648  &   $\omega\rightarrow\mathDP\pi^0$ & 0.0834941 &  $0.825 \pm 0.009$ \\
0--0.958  &    $\eta'\rightarrow\mathDP\gamma$ & 0.0219297 & $0.079 \pm 0.003$ \\
\end{tabular}
\end{table}

The branching ratios of the mesons to these new decay channels are
functions of the \mDP, the kinetic mixing parameter $\varepsilon$, the
meson type, pseudo-scalar or vector, and the meson
mass~\cite{Gorbunov:2014wqa,Batell:2009di,Berlin:2018pwi}. For
pseudo-scalar mesons $\mathcal{P}$ ($\pi^0$, $\eta^0$ and
$\eta^{\prime}$), the branching ratio to \DP$\gamma$ is given by:

\begin{align}
	\Br(\mathcal{P}\to \mathDP \gamma)\simeq 2\epsilon^2\left ( 1-\frac{\mathmDP^2}{m_{\mathcal{P}}^2} \right )^3 \Br(\mathcal{P}\to\gamma\gamma)\,. \label{eq:mesonBR1}
\end{align}
For vector mesons $\mathcal{V}$ ($\omega$), the branching ratio to a
\DP and a pseudo-scalar meson $\mathcal{P}$ is given by:

\begin{align}
\Br(\mathcal{V} \to \mathcal{P} \mathDP ) & \simeq \epsilon^2\times \Br(\mathcal{V}\to \mathcal{P} \gamma)\label{eq:mesonBR2} \\ 
&\times \frac{[(\mathmDP^2-(m_{\mathcal{V}}+m_{\mathcal{P}})^2)(\mathmDP^2-(m_{\mathcal{V}}-m_{\mathcal{P}})^2)]^{3/2}}{(m_{\mathcal{V}}^2-m_{\mathcal{P}}^2)^3}\,.\nonumber
\end{align}
For the branching ratios of the mesons to $\gamma\gamma$ or
$\gamma\pi^0$, the same values as implemented in PYTHIA 8.2 are
used. The average number of mesons produced per pp interaction,
n$_{\mathrm{meson}}$ / pp, is shown for each meson type in the last
column of Table~\ref{tab:mesonDecays}, from non-diffractive pp
collisions simulated with PYTHIA 8.2~\cite{Sjostrand:2014zea}, with
its associated statistical uncertainty.

The cross section for the production of dark photons via meson decays
produced in non-diffractive primary interactions of the proton beam is
then computed as:

\begin{align}
	\sigma_{\mathrm{meson}} = \sigma_{\mathrm{SHiP}}^{\mathrm{inel}} \times \sum_{\mathrm{mesons}}\Theta(m_{\mathrm{meson}}-\mathmDP) \times n_{\mathrm{meson}} \mathrm{/ pp} \times \Br(\mathrm{meson}\to \mathDP+X)\,,\label{eq:mesonXS}
\end{align}
using Eqs.~\eqref{eq:mesonBR1} and~\eqref{eq:mesonBR2} and values reported
in Table~\ref{tab:mesonDecays}. The
$\Theta(m_{\mathrm{meson}}-\mathmDP)$ factor is a step function
ensuring that only the mesons in the accessible mass range are
considered. To take into account the fact that the nucleon is bound in
the target, and not free as assumed by PYTHIA in our simulation of
this process, the total normalisation is taken using the inelastic
proton-nucleon cross section corresponding to the SHiP target,
$\sigma_{\mathrm{SHiP}}^{\mathrm{inel}}$ (see
Section~\ref{sec:sensitivity}). The cross section is proportional to
$\varepsilon^2$, from the dependency of $\Br(\mathrm{meson}\to
\mathDP+X)$ in Eqs.~\eqref{eq:mesonBR1} and~\eqref{eq:mesonBR2}.

\subsection{Production in proton bremsstrahlung}
\label{sec:pbrem}

In analogy with ordinary photon bremsstrahlung of scattering protons,
the same process is used for dark photon production by scattering of
the incoming 400\,GeV proton beam on the target protons. Following
Refs.~\cite{Gorbunov:2014wqa,Blumlein:2013cua}, the differential
\DP production rate can be expressed as:

\begin{align}
	\frac{\mathrm{d}^2 N}{\mathrm{d}z \mathrm{d}p_\perp^2} &= \frac{\sigma_{pp}(s')}{\sigma_{pp}(s)}\,
	w_{ba}(z,p_\perp^2)\,,\label{eq:pbremN}\\
w_{ba}(z,p_\perp^2) &= \frac{\epsilon^2 \alpha_{\mathrm{QED}}}{2\pi H}
\left[\frac{1+(1-z)^2}{z} -2z(1-z)\left(\frac{2m_p^2+\mathmDP^2}{H}
-z^2\frac{2m_p^4}{H^2}\right) \right. \nonumber \\
&\left.+2z(1-z)\big(1+(1-z)^2\big)\frac{m_p^2\mathmDP^2}{H^2}
+2z(1-z)^2\frac{\mathmDP^4}{H^2} \right]\,, \nonumber
\end{align}
where $\sigma_{pp}(s/s')$ are the total proton-proton cross sections
evaluated for the incoming/outgoing proton energy scales, $m_p$ is the
proton mass (set to $m_p = 0.938272$\,GeV~\cite{Patrignani:2016xqp}),
$P$ and $E_p$ are the proton beam initial momentum and energy
respectively, $p$ and $E_{\mathDP}$ are the momentum and energy of the
generated dark photon respectively, $p_\perp$ and $p_\parallel$ are
the components of the \DP momentum orthogonal and parallel to the
direction of the incoming proton respectively, $z$ is the fraction of
the proton momentum carried away by the dark photon in the beam
direction, $\alpha_{\mathrm{QED}}$ is the fine structure constant of
Quantum Electro Dynamic (QED), set to 1/137,
$s'=2m_p(E_p-E_{\mathDP})$, $s=2m_pE_p$ and
$H(p^2_\perp,z)=p_\perp^2+(1-z)\mathmDP^2+z^2m_p^2$.

In this formulation, the nuclear effects from having bound rather
than free protons in the target material cancel in the ratio
$\frac{\sigma_{pp}(s')}{\sigma_{pp}(s)}$.

However, the above formula does not take into account possible QCD
contributions when the mass of the emitted \DP exceeds that of the
proton, and the bremsstrahlung process starts to depend on the
internal partons. It does not take into account the possibility of
enhancement in the cross section due to nuclear resonances in the
so-called vector meson dominance (VMD) model either. In consequence,
two independent approaches are followed, leading to two different
estimates of the final cross section.

In the first approach, when the mass of the dark photon is larger than
1\,GeV, the standard dipole form factor~\cite{RevModPhys.35.335} is
included in the proton-\DP vertex, leading to a penalty factor that
models the strong suppression of the bremsstrahlung production:

\begin{align}
	\mathrm{penalty}(\mathmDP) = {\left(\frac{\mathmDP^2}{0.71\,\mathrm{GeV}^2}\right)}^{-4}\quad \mathrm{ for }\quad \mathmDP^2 > 0.71\,\mathrm{GeV}^2\,.
\end{align}
According to Ref.~\cite{Gorbunov:2014wqa}, this form factor is
conservative and probably underestimates the rates. The direct
parton-parton QCD production will dominate above 1.5\,GeV and is
described in Section~\ref{sec:qcd}.

In the second approach, the VMD form factor from
Refs.~\cite{deNiverville:2016rqh,Faessler:2009tn} is used, leading to
an enhancement of the cross section by a factor $10^4$ around the
$\rho$ and $\omega$ meson mass of 0.8\,GeV, and still up to a factor
10 in the tail due to also considering resonances of masses 1.25 and
1.45\,GeV following the description in Ref.~\cite{Faessler:2009tn}.

The total p-p cross section $\sigma_{pp}(s)$ is taken
from experimental data:

\begin{align}
\sigma_{pp}(s) = Z + B\cdot \log^2\left(\frac{s}{s_0}\right) + Y_1\left(\frac{s_1}{s}\right)^{\eta_1} - 
Y_2\left(\frac{s_1}{s}\right)^{\eta_2},
\end{align}
where $Z = 35.45$~mb, $B = 0.308$~mb, $Y_1 = 42.53$~mb, $Y_2 =
33.34$~mb, $\sqrt{s_0} = 5.38$~GeV, $\sqrt{s_1} = 1$~GeV, $\eta_1 =
0.458$ and $\eta_2 = 0.545$~\cite{Nakamura_2010}. This formulation has
been compared to the latest parameterisation from
Ref.~\cite{Patrignani:2016xqp}, 
and found to be almost identical for the momentum range of interest
here.

\begin{figure}[h!]
  \centering
\includegraphics[width=0.48\textwidth]{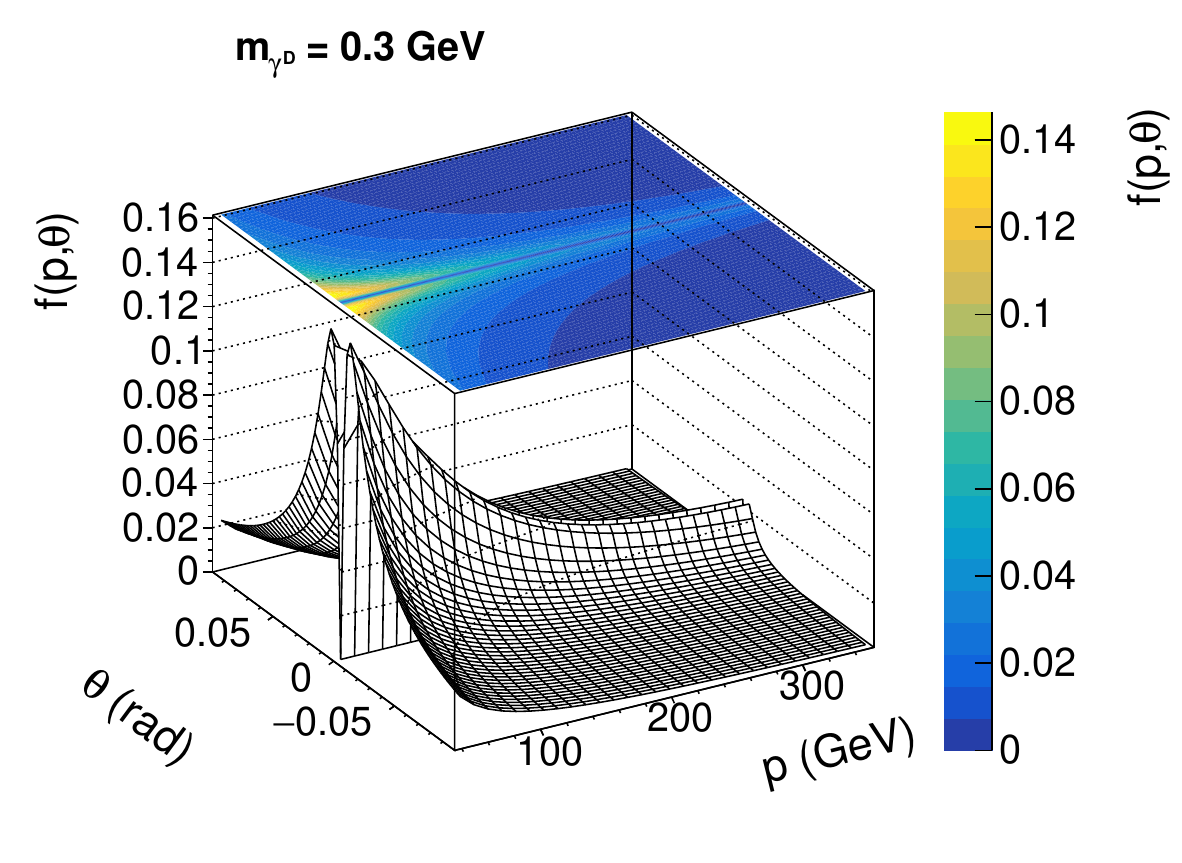}
\hfill
\includegraphics[width=0.48\textwidth]{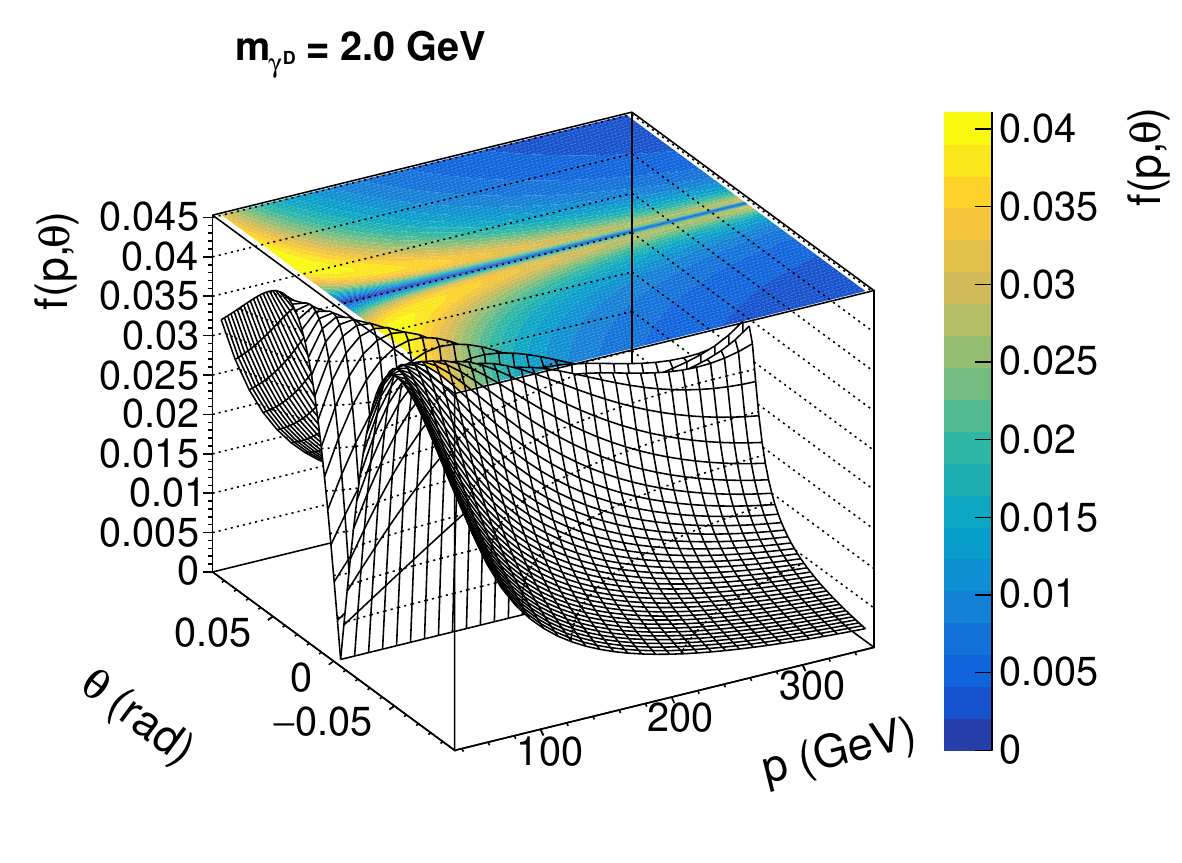}
\caption{Normalised probability density function of producing a dark
  photon with angle $\theta$ and momentum $p$ through proton
  bremsstrahlung, for two representative examples of \mDP: 0.3\,GeV
  (left) and 2\,GeV (right).}
\label{fig:pbremPDF}
\end{figure}

Reformulating Eq.~\eqref{eq:pbremN} as a function of the \DP angle
$\theta$ to the beam line and its total momentum p, a two-dimensional
normalised probability density function (PDF) $f(p, \theta)$
is extracted, and shown in Fig.~\ref{fig:pbremPDF} for two
representative choices of \mDP. Note that due to the simple dependency
of the production rate scaling as $\varepsilon^2$, the normalised PDF
is independent of $\varepsilon$. The dark photons are generated with
maximum probability on each side of the beam axis ($\theta$ close to
0) with a factor of 5 more chance to have $p<100$\,GeV
compared to $p>200$\,GeV, for the low masses, and increased
probability to have high momentum as the mass increases.

Events are generated using a PYTHIA 8 particle gun with the \DP as
particle, randomly choosing the \DP(p, $\theta$) values according to
the normalised 2D PDF $f(p, \theta)$, extracted for each \mDP point
studied.

The integral of $\frac{\mathrm{d}^2 N}{\mathrm{d}p \mathrm{d}\theta}\times\mathrm{FF}$, with FF
the penalty dipole form factor or the VMD form factor, in the range of
momenta and solid angle kinematically allowed, provides an estimate of
the dark photon production rate per pp interaction through proton
bremsstrahlung, scaling as $\varepsilon^2$. The production cross
sections using the dipole form factor and VMD form factor methods are
expressed by:

\begin{align}
	\sigma_{\mathrm{pbrem}} = \sigma_{\mathrm{SHiP}}^{\mathrm{inel}} \times \int_{p_{\mathrm{min}}}^{p_{\mathrm{max}}}{\int_{\theta=-\theta_{\mathrm{max}}}^{\theta_{\mathrm{max}}}{ \mathrm{FF} \times \frac{\mathrm{d}^2 N}{\mathrm{d}p \mathrm{d}\theta}\,\mathrm{d}\theta}\mathrm{d}p}\,, \label{eq:pbremXS}
\end{align}
and shown in Fig.~\ref{fig:pbremxs}. The conditions of validity of the
approximation used to derive
Eq.~\eqref{eq:pbremN}~\cite{Kim_Tsai_1973,KIM1972665} require a lower
momentum bound for the \DP at $p_{\mathrm{min}}=0.1
P_{\mathrm{p}}$~\cite{Blumlein:2013cua}, and an upper bound at
$p_{\mathrm{max}}=0.9 P_{\mathrm{p}}$, as well as an upper bound on
$p_\perp<4$\,GeV, giving $\theta_{\mathrm{max}} \simeq 0.1$ rad.

\subsection{Drell-Yan production}
\label{sec:qcd}

For production of the dark photon in parton-parton scattering, the
generic implementation of a resonance that couples both to SM fermion
pairs and hidden particles is used, as implemented in PYTHIA 8.2 under
the ``HiddenValley'' Z$^{\prime}$ model~\cite{Carloni:2011kk}. A cross-check
has been done that similar kinematic distributions for the dark
photons are found using another Z$^{\prime}$ implementation in PYTHIA from
the ``New Gauge Bosons'' class of processes~\cite{Ciobanu:2005pv}.

The dark photons are generated in the mass range $1.4 < \mathmDP\ <
10$\,GeV. Below 1.4\,GeV one leaves the domain of perturbative QCD and
the parton model cannot be used anymore.

\begin{figure}[h!]
  \begin{minipage}[b]{0.48\textwidth}
    \includegraphics[width=1.\textwidth]{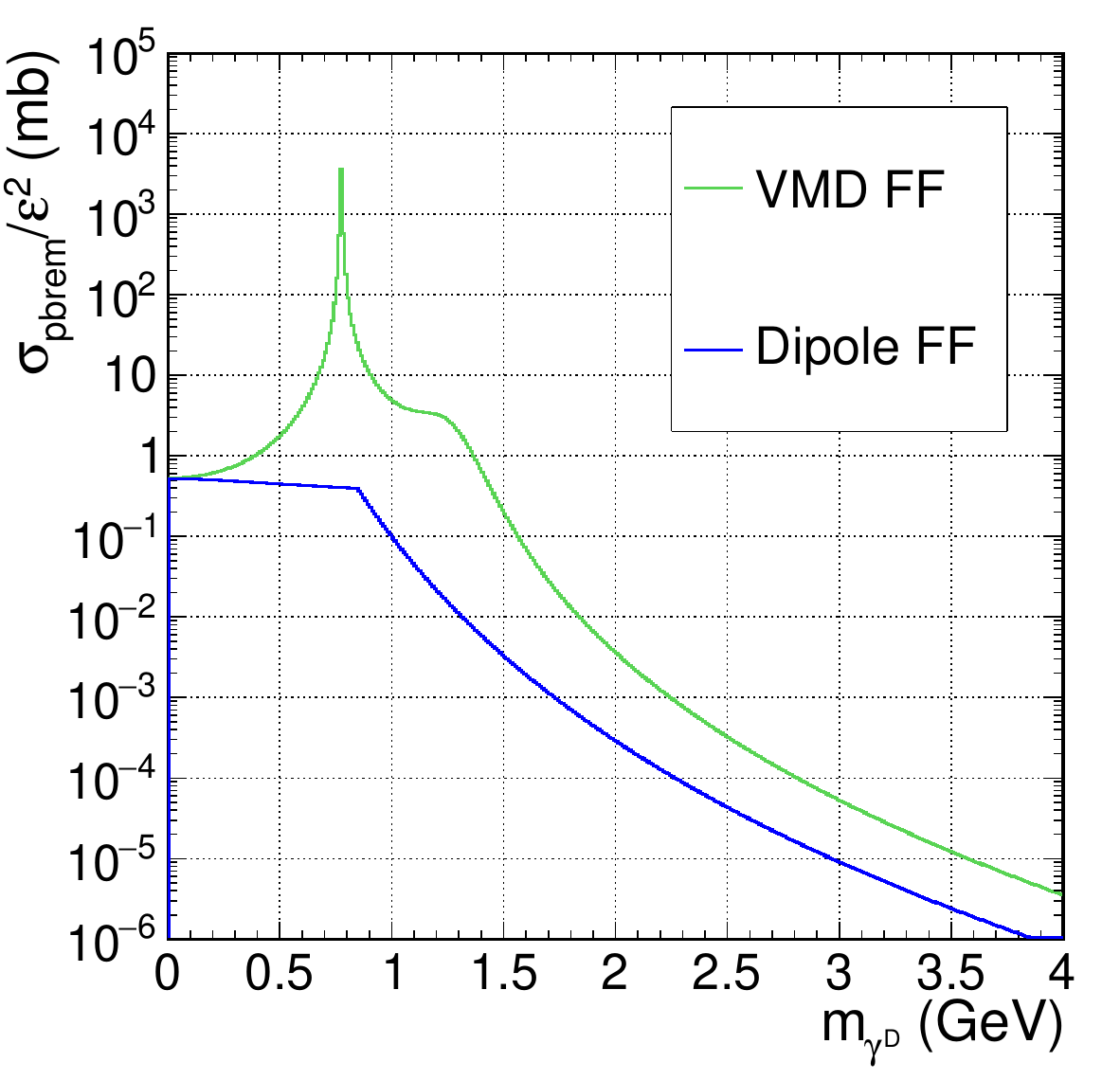}
    \caption{Proton bremsstrahlung production cross section as a function of \mDP.}
    \label{fig:pbremxs}
  \end{minipage}
  \hfill
  \begin{minipage}[b]{0.48\textwidth}
    \includegraphics[width=1.\textwidth]{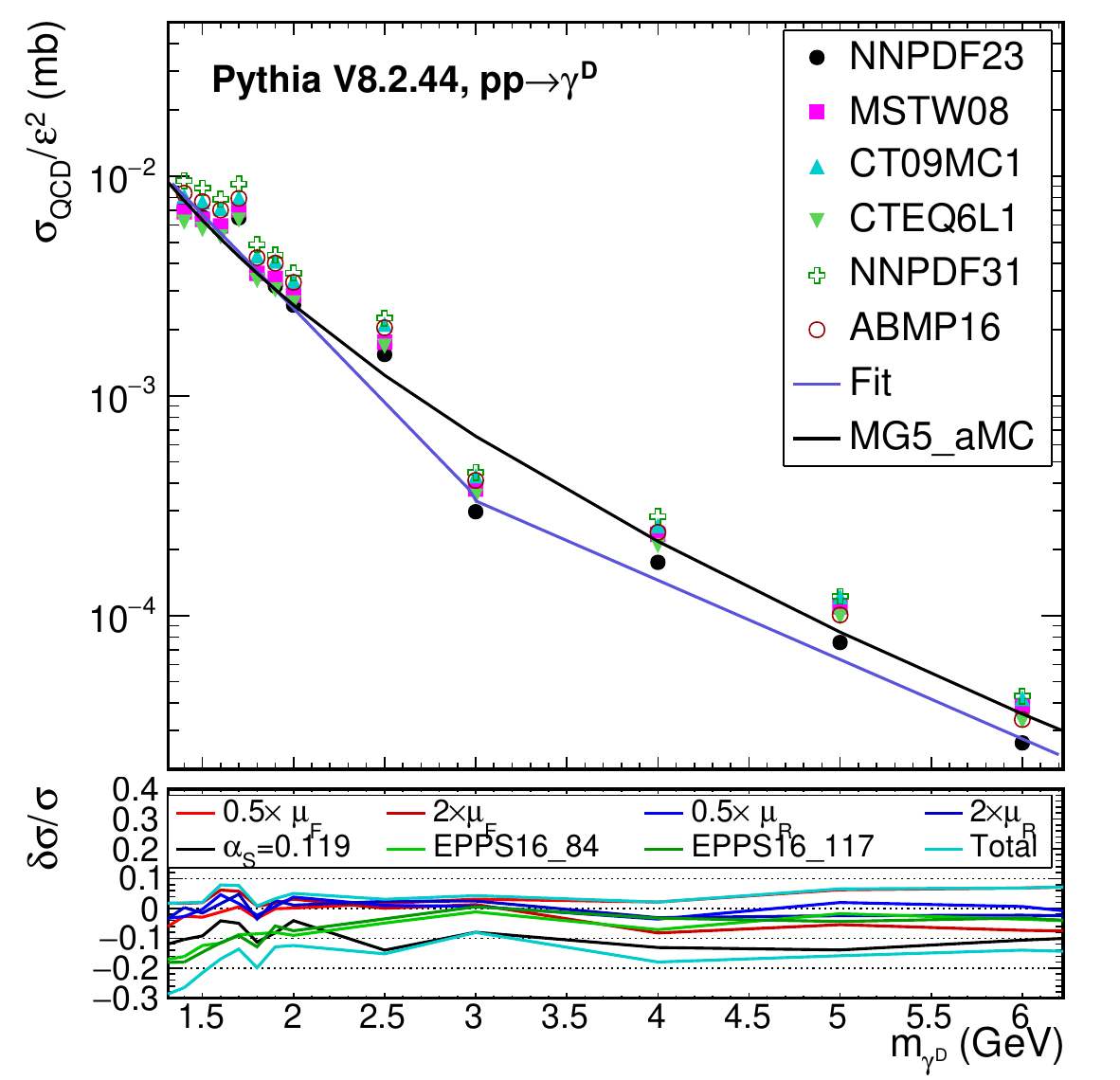}
    \caption{QCD production cross section at LO as a function of \mDP. The
      fit function is described in Eq.~\eqref{eq:qcdXS}. The upper pad
      shows the impact of different PDF
      sets~\cite{Ball:2017nwa,Martin:2009iq,Lai:2009ne,Pumplin:2002vw,Alekhin:2017kpj}. The
      lower pad shows the relative uncertainties from several
      theoretical uncertainty sources on the cross section calculated
      by PYTHIA, and their sum in quadrature under ``Total''.}
    \label{fig:qcdXS}
  \end{minipage}
\end{figure}

The LO cross section given by PYTHIA when the new particle has the
properties of the dark photon is shown in Fig.~\ref{fig:qcdXS}. The
nuclear effects are neglected and the parton-parton cross section from
PYTHIA is used without modification. Like for the meson and proton
bremsstrahlung processes, it is found to scale as $\varepsilon^2$. The
LO NNPDF2.3 PDF set~\cite{Ball:2012cx} has been used with the default
Monash 2013 tune~\cite{Skands:2014pea}, and the strong coupling
constant set to $\alpha_{\mathrm{s}} = 0.13$. An empirical function is
extracted to parameterise the cross section (in mb) as a function of
the \DP mass (in GeV) in a continuous way, described in
Eq.~\eqref{eq:qcdXS}. The impact of several sources of theoretical
uncertainties (PDF choice~\cite{Buckley:2014ana}, QCD scales,
$\alpha_{\mathrm{s}}$) are studied and shown in Fig.~\ref{fig:qcdXS}
(see also Section~\ref{sec:systs}). The impact of nuclear effects is
checked using the nuclear modification factors available in PYTHIA,
with the most recent nuclear PDF set EPPS16~\cite{Eskola:2016oht},
using the two atomic masses ($\mathcal{A}=84$ and $117$) available
around the SHiP target material one ($\mathcal{A}=96$). Both give very
similar results, with a cross section varying within $\pm 6$\% from
the NNPDF2.3 proton PDF one, depending on the \DP mass. The
alternative generator Madgraph5\_aMC@NLO v2.7.2~\cite{Alwall:2014hca}
is also used to cross check the cross section calculation. The
parameterisation of the width of the resonance, dependent on the
branching ratios to fermion pairs, is a little different and explains
the difference seen.

\begin{align}
  1.4 < \mathmDP\ \leq 3\,\mathrm{GeV} & : \sigma_{\mathrm{QCD}} =  \varepsilon^2 \times e^{-2.05488-1.96804\times \mathmDP}\,, \label{eq:qcdXS} \\
  \mathmDP\ > 3\,\mathrm{GeV} & : \sigma_{\mathrm{QCD}} =  \varepsilon^2 \times e^{-5.51532-0.830917\times \mathmDP}\,. \nonumber
\end{align}
Higher-order contributions to the process could lead to a sizable
increase of the cross section at such low masses. Using the MATRIX
v1.0.5
program~\cite{Grazzini:2017mhc,Catani:2009sm,Cascioli:2011va,Denner:2016kdg,Catani:2012qa,Catani:2007vq},
the ratio of next-to-next-to-leading-order (NNLO) over LO differential
cross sections for standard Drell-Yan production at
$\sqrt{s}=27.43$\,GeV as a function of the dilepton invariant mass
M$_{\ell\ell}$ is found to be rather flat at $1.7 \pm 0.17$\,(stat)
for M$_{\ell\ell}$ between 1.4 and 5\,GeV, increasing up to 2.2 at
M$_{\ell\ell}=10$\,GeV. The QCD scale uncertainties on the ratio are
found to be $_{-11\%}^{+20\%}$ in the range $2 < $M$_{\ell\ell} <
5$\,GeV increasing above $\pm 30\%$ for M$_{\ell\ell}$ above 10\,GeV
or below 2\,GeV. The MSTW2008 NNLO PDF set~\cite{Martin:2009iq} has
been used for all calculations. These calculations are also found to
be in fair agreement (within 10\%) with a study performed more
specifically in proton-antiproton collisions, with a special interest
for low dilepton masses and in particular for $\sqrt{s}=30$\,GeV, from
Ref.~\cite{Shimizu:2005fp}, and considering soft-gluon resummations at
all orders. Given the lack of experimental data at these low masses
and low $\sqrt{s}$ to confirm the size of the expected correction and
the impact from PDF and non-perturbative effects on the actual dark
photon production, a final k-factor of $1.7 \pm 0.7$ is applied to the
LO PYTHIA cross section from Eq.~\eqref{eq:qcdXS}.

\begin{figure}[h!]
  \centering
  \includegraphics[width=0.48\textwidth]{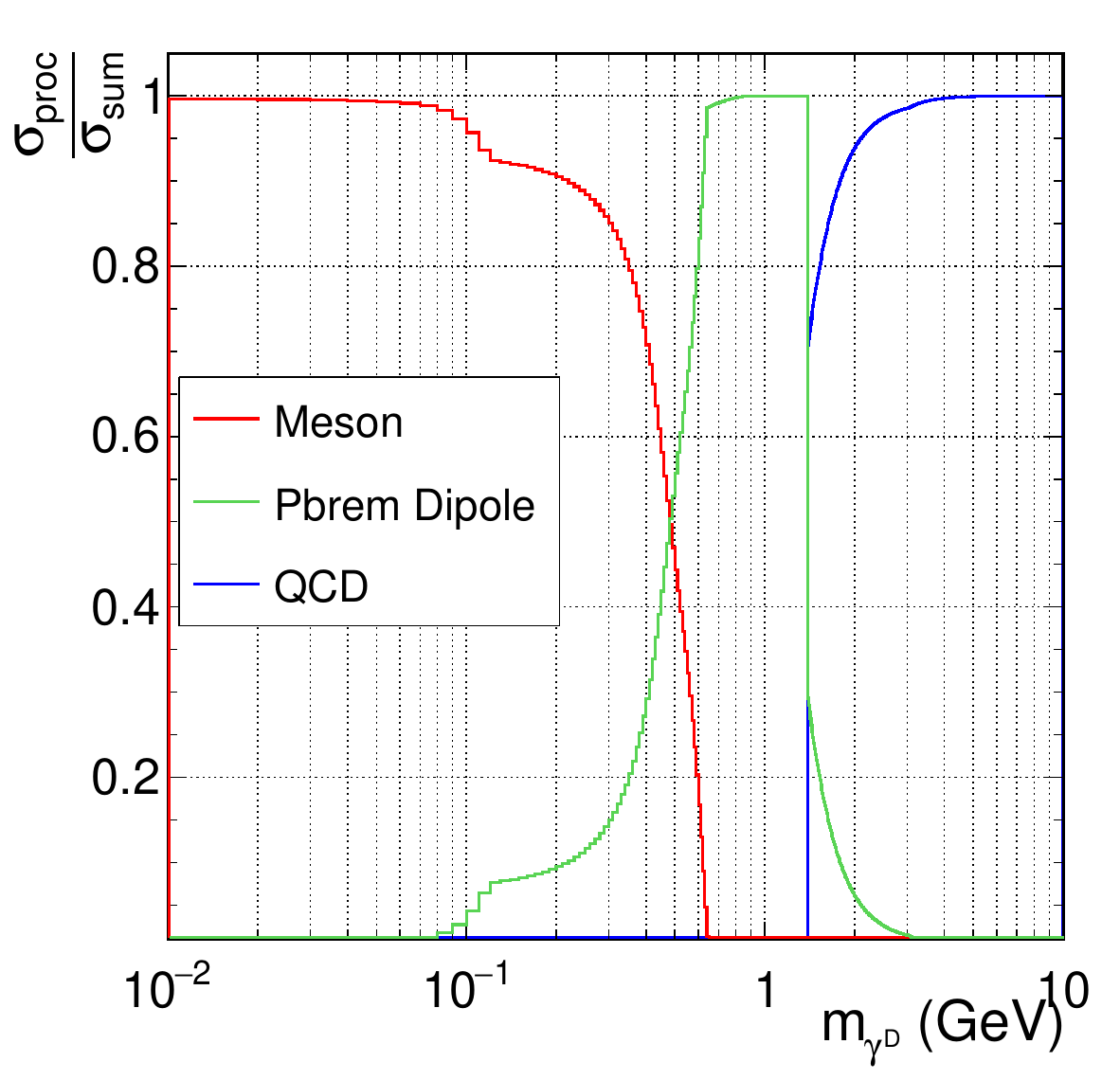}
  \hfill
\includegraphics[width=0.48\textwidth]{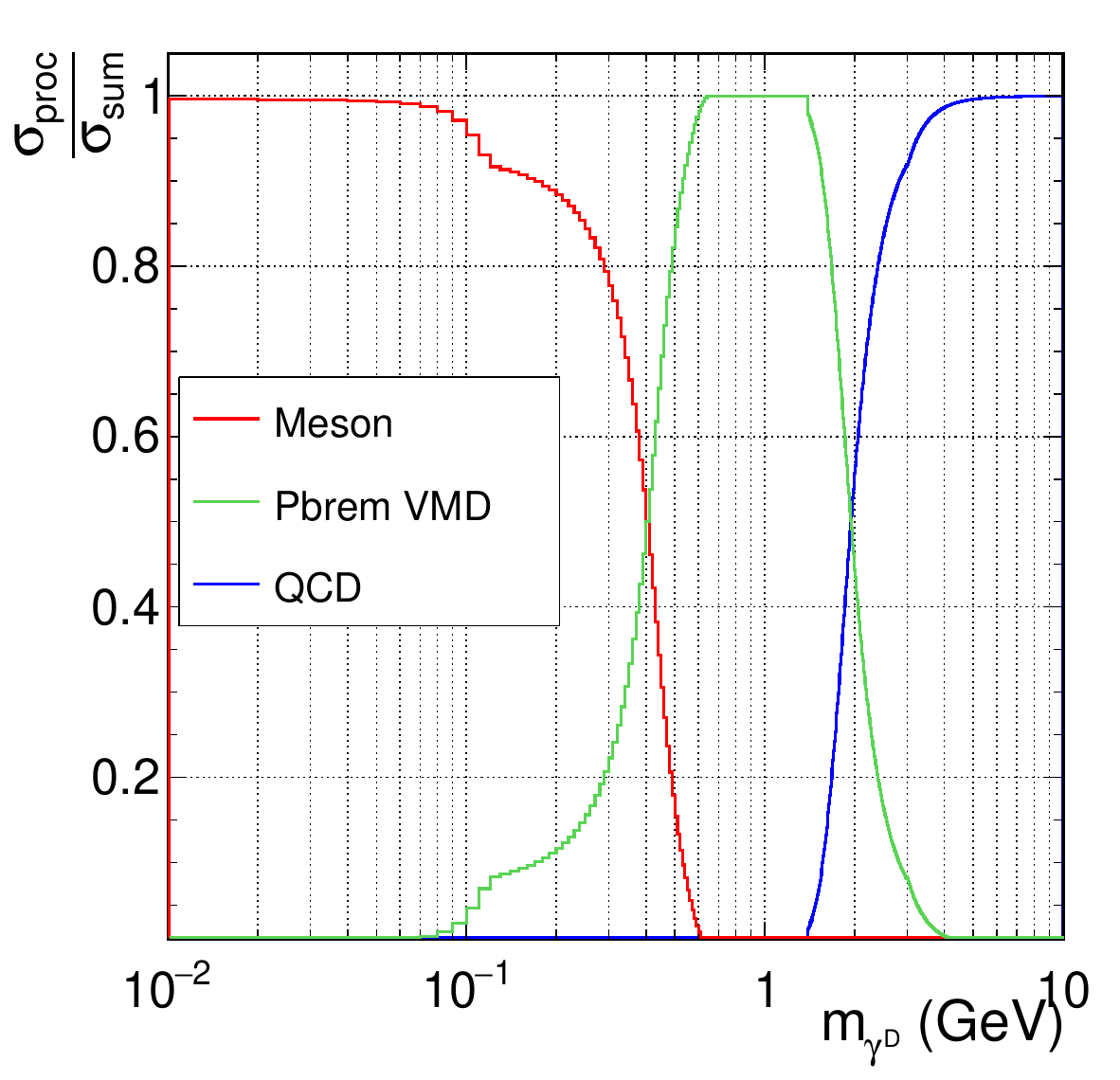}
\caption{Relative contributions to the cross section as a function of \mDP for the three production modes studied, using the dipole form factor for proton bremsstrahlung (left) or the VMD form factor (right).}
\label{fig:allXS}
\end{figure}

The relative contribution from each process is shown in
Fig.~\ref{fig:allXS}, as a function of \mDP, for the three
production modes, in the two scenarios considered for the proton
bremsstrahlung mode.

\subsection{Dark photon decays}
\label{sec:decay}

Except for the meson production mode, in which the new particle
couples to the parent meson via mixing with the photon and hence
cannot be a resonance from PYTHIA's point-of-view, in QCD and proton
bremsstrahlung the \DP is implemented as a resonance. In all cases,
the decay channels are implemented as follows.

The partial decay width of the dark photon into a lepton pair is given
by~\cite{Blumlein:2013cua}:

\begin{align}
	\Gamma(\mathDP \rightarrow \ell^+ \ell^-) = \frac{1}{3} \alpha_{\mathrm QED} \mathmDP \epsilon^2 \sqrt{1 - \frac{4 m_{\ell}^2}{\mathmDP^2}}\left(1 + \frac{2 m_{\ell}^2}{\mathmDP^2}\right)~,\label{eq:dp-decay}
\end{align}
where $m_{\ell}$ is the lepton mass, for electron, muon or tau
leptons, if kinematically allowed. Following the approach used by the
authors of Ref.~\cite{Bjorken:2009mm}, the partial decay width into
quark pairs is computed as:

\begin{align}
	\Gamma(\mathDP \rightarrow~\text{hadrons}) = \Gamma(\mathDP \rightarrow \mu^+ \mu^-) 
	R\left(\mathmDP\right)~,
\end{align}
where

\begin{align}
	R\left(\sqrt{s}\right) = \frac{\sigma(e^+e^- \rightarrow~\text{hadrons})}{\sigma(e^+e^- \rightarrow \mu^+\mu^-)}\,,
\end{align}
is the energy-dependent R-ratio quantifying the hadronic annihilation
in $e^+e^-$ collisions~\cite{Agashe:2014kda}, tabulated from 0.3 to
10.29\,GeV.

The lifetime of the \DP is then naturally set to the inverse of its
total width, summing all the kinematically-allowed channels for
calculating the total width. It is proportional to
1/$\varepsilon^2$. The branching ratios to individual channels are set
to the ratio of the partial over total width, and are hence
independent of $\varepsilon$. For separating the hadronic channels
into the different quark-flavoured pairs allowed kinematically, the
coupling is assumed to be proportional to the quark charge $q$ as
$\mathrm{n}_{\mathcal{C}} \times q^2$~\cite{Liu:2017ryd}, with
$\mathrm{n}_{\mathcal{C}}=3$ the number of coloured charges. When the
\DP is implemented as a resonance in PYTHIA, the decay goes explicitly
through the pair of quarks, before hadronisation. Otherwise the
hadrons are found as direct decay products of the \DP.

\begin{figure}[h!]
\begin{minipage}[b]{0.48\textwidth}
\includegraphics[width=1.\textwidth]{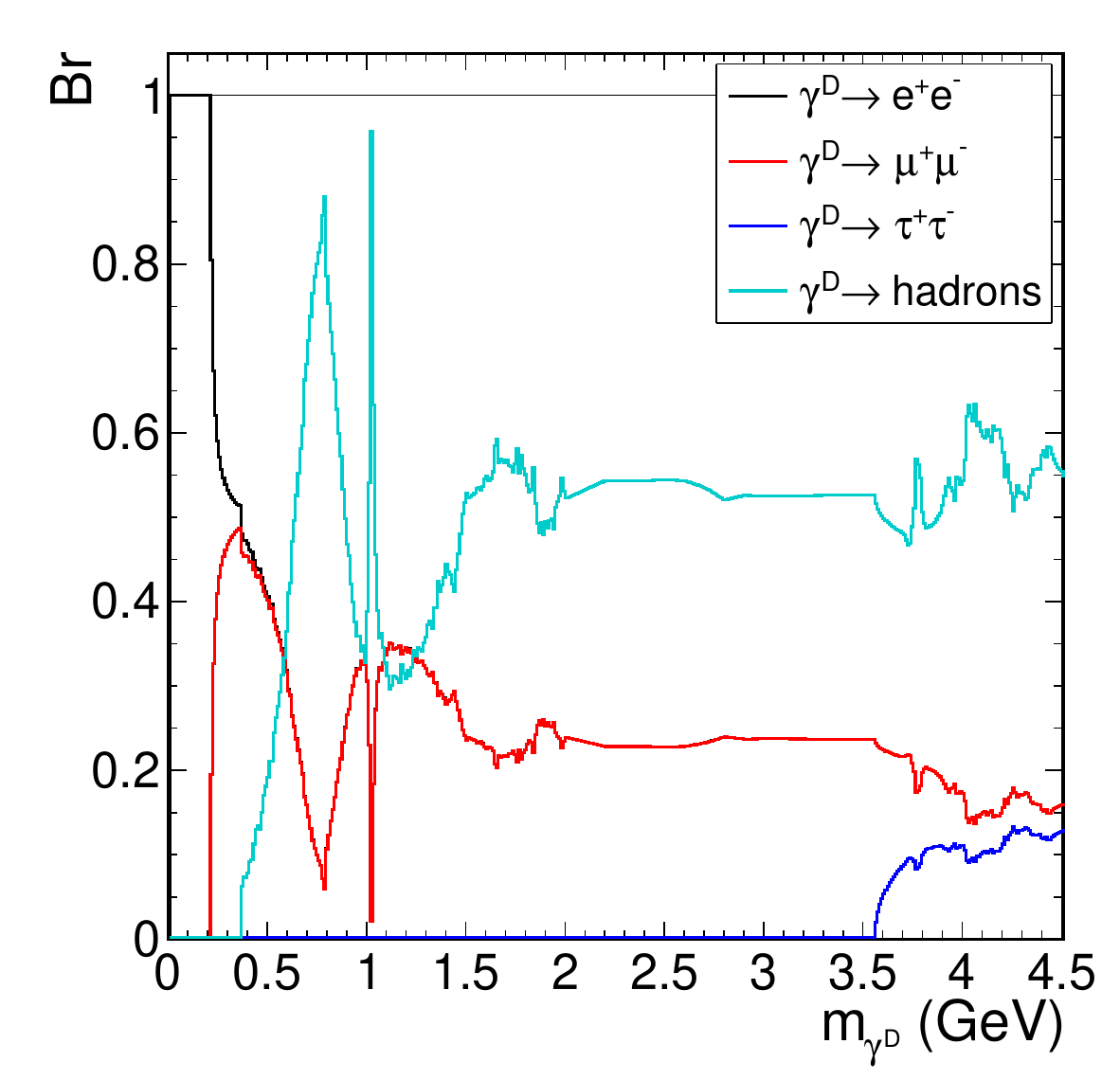}
\caption{Branching ratio of the \DP into pairs of leptons or quarks as
  a function of its mass.}
\label{fig:decayBR}
\end{minipage}
\hfill
\begin{minipage}[b]{0.48\textwidth}
  \includegraphics[width=1.\textwidth]{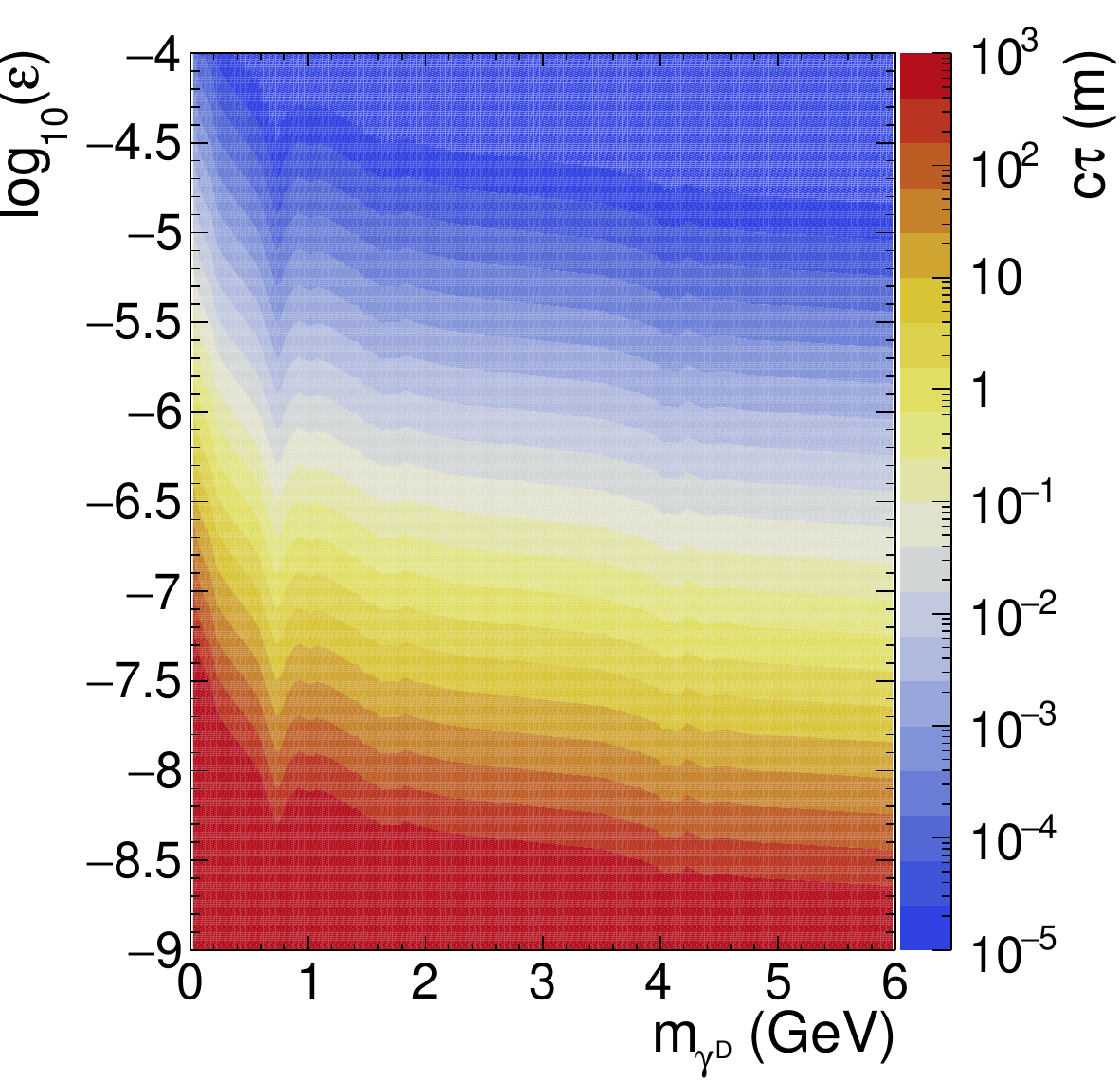}
  \caption{Expected lifetime of the dark photon as a function of its mass and of the kinetic mixing parameter $\varepsilon$.}
  \label{fig:ctau}
\end{minipage}
\end{figure}

The branching ratio of the \DP into pairs of leptons or quarks is
shown in Fig.~\ref{fig:decayBR} as a function of \mDP. The hadronic
decays become available above the pion mass threshold. The expected
lifetime of the \DP as a function of its mass and $\varepsilon$ mixing
parameter is shown in Fig.~\ref{fig:ctau}.

\section{SHiP sensitivity}
\label{sec:sensitivity}

In order to maximise the statistical power of the limited number of
events produced with PYTHIA in the different production modes, the \DP
decay vertex position is randomly assigned to be inside the decay
vessel of length L$_{\mathrm{Vessel}}= 50.760$\,m, and the associated
probability of this happening is given as a function of the \DP
four-momentum (p, E$_{\mathDP}$) and lifetime $\mathrm{c}\tau$:

\begin{align}
  \mathrm{w}_{\mathrm{vtx}}(\ell) = e^{-\frac{\ell+\mathrm{L}_0}{\beta\times\gamma\times \mathrm{c}\tau}}\times \frac{\mathrm{L}_{\mathrm{Vessel}}}{\beta\times \gamma \times \mathrm{c}\tau}\,,\label{eq:decayVtx}
\end{align}
with $\gamma = E_{\mathDP}/\sqrt{E_{\mathDP}^2-p^2}$, $\beta =
p/E_{\mathDP}$, L$_{0}$ is the distance from the target to the
entrance of the decay vessel, and $\ell$ is randomly distributed
between 0 and L$_{\mathrm{Vessel}}$ with a flat prior.

The total event rate expected is then extracted from the cross
sections $\sigma_{\mathrm{prod}}$ defined in Section~\ref{sec:DPprod},
i.e. Eqs.~\eqref{eq:mesonXS},~\eqref{eq:pbremXS} and~\eqref{eq:qcdXS}
for the meson, proton bremsstrahlung and QCD productions,
respectively, normalising to the luminosity expected from the
$\mathrm{N} = 2 \times 10^{20}$ proton-on-target events that
will be collected by the end of the SHiP physics program. The expected
rate is taking into account the detector acceptance and the efficiency
to reconstruct the decay products in the SHiP detector,
$\mathcal{P}_{\mathrm{vessel}}$ and $\mathcal{P}_{\mathrm{reco}}$
described in detail in Sections~\ref{sec:vtx} and~\ref{sec:reco},
respectively, and following Eq.~\eqref{eq:finalN}:

\begin{align}
\mathcal{N}_{\mathDP} = \sigma_{\mathrm{prod}} \times \mathcal{L}_{\mathrm{SHiP}} \times \Br(\mathDP \to \mathrm{ch}+\mathrm{ch}) \times \mathcal{P}_{\mathrm{vessel}} \times \mathcal{P}_{\mathrm{reco}}\,.  \label{eq:finalN}
\end{align}
The SHiP luminosity is defined as:
$\mathcal{L}_{\mathrm{SHiP}}=\frac{\mathrm{N}}{\sigma_{\mathrm{SHiP}}^{\mathrm{inel}}}$,
using an inelastic proton-nucleon cross section of
$\sigma_{\mathrm{SHiP}}^{\mathrm{inel}}=10.7$\,mb~\cite{Anelli:2007512}, which
directly corresponds to the SHiP target material (Molybdenum) nuclear
interaction length and density.

The strategy of the analysis relies on identifying the decays of the
\DP\ into at least two charged particles, $\mathDP \to
\mathrm{ch}+\mathrm{ch}$. The reconstructed charged tracks must
originate from a common vertex. These requirements are enough to
ensure that almost no background event will survive the selection, as
demonstrated in Refs.~\cite{Alekhin_2016,Anelli:2007512}. The 90\% confidence level (CL) limits
on the existence of a \DP with given (\mDP, $\varepsilon$) are hence
set by excluding regions where more than $\mathcal{N}_{\mathDP} =2.3$ events are expected.

\subsection{Decay channels}

\begin{figure}[h!]
  \centering
\includegraphics[width=0.6\textwidth]{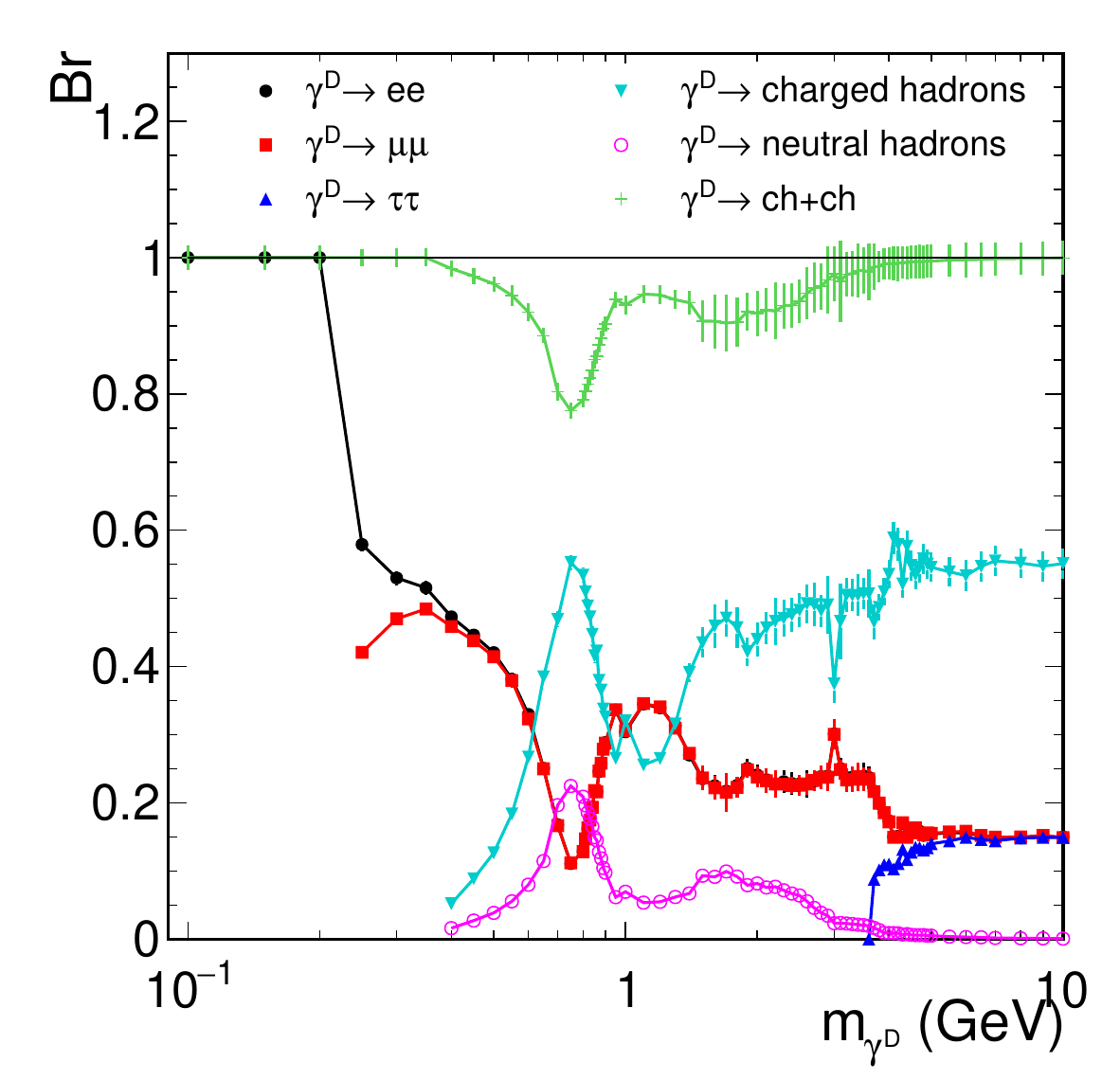}
\caption{Branching ratio to the visible decay channels, as a function
  of \mDP. $\Br(\mathDP \to \mathrm{ch}+\mathrm{ch})$ is equal to
  $1-\Br(\mathDP \to$ neutral hadrons).}
\label{fig:brdecay}
\end{figure}

The following final states are considered, whenever available for a
given \mDP: e$^{+}$e$^{-}$, $\mu^{+}\mu^{-}$, $\tau^{+}\tau^{-}$, and
any hadronic decay channels leading to charged particles
(e.g. $\pi^{+}\pi^{-}+X$, $K^{+}K^{-}+X$). The branching ratio to the
different final states is shown in Fig.~\ref{fig:brdecay} for all the
simulated (\mDP, $\varepsilon$) points in the three different
production modes, as a function of \mDP, calculating the mean value
over the different $\varepsilon$ samples. All events classified under
``e$^{+}$e$^{-}$'', ``$\mu^{+}\mu^{-}$'', ``$\tau^{+}\tau^{-}$'' and
``charged hadrons'' have at least two charged particles, their sum is
represented as ``$\mathrm{ch}+\mathrm{ch}$''. Only the events
classified under ``neutral hadrons'' are lost due to the analysis
selection described in Section~\ref{sec:reco}. Compared to
Fig.~\ref{fig:decayBR}, Fig.~\ref{fig:brdecay} highlights the mass
scan actually simulated, and the separation of the hadronic final
states into the charged and neutral ones.

\subsection{Vessel acceptance}
\label{sec:vtx}

\begin{figure}[h!]
  \centering
\includegraphics[width=1.\textwidth]{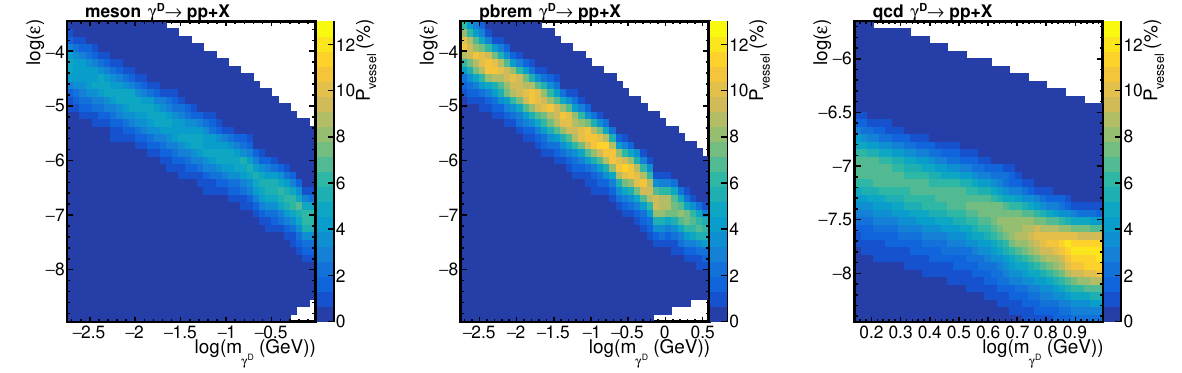}
\caption{Efficiency of requiring the \DP decay vertex to be inside the
  decay vessel volume, for the three production modes studied. An
  interpolation between the simulated (\mDP, $\varepsilon$) values is
  performed using a linear interpolation via Delaunay triangulation.}
\label{fig:Pvessel}
\end{figure}

For events which have two charged particles, the \DP decay vertex is
further required to be within the vessel volume. The efficiency of
this selection, $\mathcal{P}_{\mathrm{vessel}}$ is defined as the
ratio of the sum of the weights $\mathrm{w}_{\mathrm{vtx}}(\ell)$ of
events passing the vertex selection described in
Table~\ref{tab:recoSel} over the total number of events with a dark
photon decaying to at least two charged particles. This efficiency is
shown in Fig.~\ref{fig:Pvessel} as a function of (\mDP,
$\varepsilon$), for the three production modes. It is mostly driven by
the lifetime of the \DP, and the kinematics of the \DP produced in the
target. Its maximum is around 5\% for the production via meson decay,
10\% for the proton bremsstrahlung production, and for higher masses
in QCD production.

\subsection{Reconstruction of the decay products}
\label{sec:reco}

The strategy employed in this analysis relies uniquely on the
reconstruction of charged particles by the SHiP straw tracker. Future
extensions of this work could consider also calorimeter deposits (with
the possibility to fully reconstruct $\pi^0$ decays to two photons)
and muon detectors. Events are retained if two tracks are found
passing the criteria summarised in Table~\ref{tab:recoSel}, namely
that the two tracks are within the fiducial area of the detector up to
the fourth layer after the magnet, the fit converged with good quality
requirements ($\chi^2/\mathrm{NDF}<5$ with $\mathrm{NDF}$ the number
of degrees of freedom of the fit). The tracks are required to have an
impact parameter (IP) less than 0.1\,m in the (x,y) plane, a momentum
$p$ above 1\,GeV, and a distance of closest approach (DOCA) below
1\,cm. Criteria on the number of hits ($\mathrm{NDF}>25$) or presence
of hits before/after the magnet are meant to reduce backgrounds which
could come from particles re-entering the detector volume due to the
magnetic field. At the moment, the resolution of the timing detector
is neglected, and MC truth information is used instead.

\begin{table}[thbp]
  \caption{Selection criteria applied on the reconstructed events. See Fig.~\ref{fig:stations} for the layout simulated.}
  \label{tab:recoSel}
  \centering
  \begin{tabular}{p{0.2\textwidth}|p{0.8\textwidth}}
    Decay vertex & z position within the range [610,5076] cm.\\
    & x-y within vessel volume and at least 5 cm away from its inner walls.\\
    \hline
    Straw tracker hits & in each layer - before and after magnet - up to tracking station 4 \\
    \hline
    Tracks & $\geq 2$ tracks \\
    & $\mathrm{NDF} > 25$, $\chi^2 / \mathrm{NDF} < 5$, DOCA $<1$\,cm, p$>1$\,GeV, IP$<0.1$\,m \\
\end{tabular}
\end{table}

\begin{figure}[h!]
  \centering
\includegraphics[width=1.\textwidth]{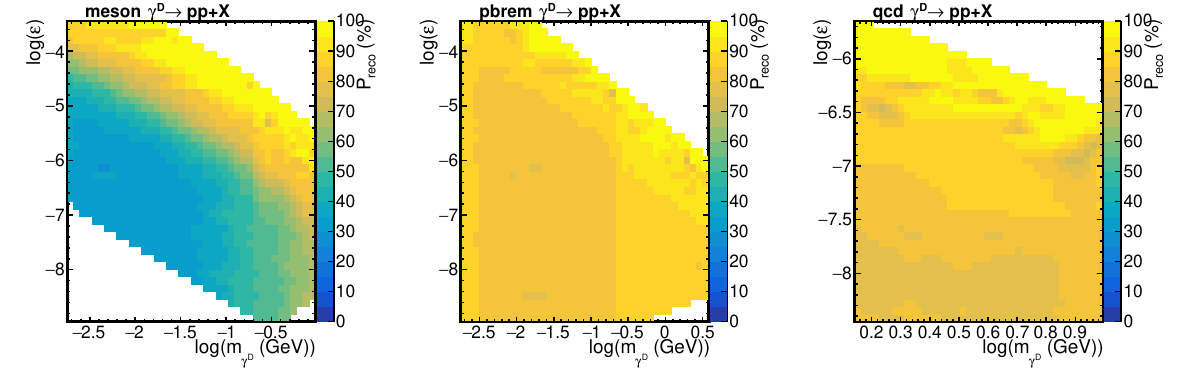}
\caption{Efficiency of requiring two good tracks, for events with two
  charged particles and \DP vertex inside the vessel volume, for the
  three production modes studied. An interpolation between the
  simulated (\mDP, $\varepsilon$) values is performed using a linear
  interpolation via Delaunay triangulation.}
\label{fig:Preco}
\end{figure}

The efficiency of having two good tracks passing the selection for
events which had two charged particles and \DP vertex in the decay
volume, $\mathcal{P}_{\mathrm{reco}}$ is shown in
Fig.~\ref{fig:Preco}. Once the \DP decays in the vessel volume, the
reconstruction efficiency is above 80\% in most of the parameter
space. For production via meson decay, a dependency on $\varepsilon$
is observed, with the efficiency dropping to below 50\% as
$\varepsilon$ decreases. This is found to be related to the wider
angular distribution of dark photons produced in meson decays,
introducing a dependency on the position of the decay vertex.

\subsection{Systematic uncertainties}
\label{sec:systs}

The following sources of systematic uncertainties from theory are
investigated, for the three production modes. The missing
contributions from cascade decays will be the subject of future work
and is not considered.

For the meson production, the overall rate is affected by the
following uncertainties:
\begin{itemize}
\item Branching ratios of the mesons to decay channels with photons from
  Table~\ref{tab:mesonDecays}: from Ref.~\cite{Patrignani:2016xqp}, the
  uncertainties on the measurement of these branching ratios are
  0.03\%, 0.5\%, 3.4\% and 3.6\% for $\pi^0\rightarrow\gamma\gamma$,
  $\eta^0\rightarrow \gamma\gamma$, $\omega\rightarrow \pi^{0}\gamma$
  and $\eta^{\prime}\rightarrow \gamma\gamma$ respectively,
  translating directly to the final rate.
\item Uncertainty on the meson multiplicities and shape of their
  kinematics properties: PYTHIA 8.2 has been compared with data in
  several existing publications. In Ref.~\cite{Dobrich:2019dxc}, a
  comparison to NA27 and NA56 data is made for the inclusive
  production of $\pi^0$ mesons, and reasonable agreement is found,
  within 30\% in the kinematic regions targeted by our
  measurement. PHENIX and ALICE also measured inclusive $\pi^0$,
  $\eta$ and $\omega$ production and
  ratios~\cite{Acharya:2017tlv,ALICE-PUBLIC-2018-004,Adare:2011ht},
  and showed global agreement within about 20\% with the PYTHIA 8
  (Monash 2013 Tune) simulation.
\end{itemize}
Adding the different sources in quadrature, this results in a total
systematic uncertainty of $\pm 30 \%$.

For the proton bremsstrahlung, the theory systematic uncertainties concern:
\begin{itemize}
\item uncertainties on the inelastic p-p cross section
  $\sigma_{pp}(s)$, which will mostly cancel in the ratio
  $\frac{\sigma_{pp}(s')}{\sigma_{pp}(s)}$, are neglected.
\item Dipole form factor versus VMD form factor: the two scenarios are presented separately in the final exclusion limits.
\item Contribution from protons undergoing elastic scattering before
  radiating the \DP: an upper bound is derived using a factor
  $\frac{1}{1-P_{\mathrm{el}}} =
  1.34$~\cite{shipLDM}, with
  $P_{\mathrm{el}}$ the probability for an incoming proton to generate
  an elastic scattering, $P_{\mathrm{el}} =
  \frac{\sigma_{\mathrm{pp}}^{\mathrm{elastic}}}{\sigma_{\mathrm{pp}}^{\mathrm{tot}}}$
  and $\sigma_{\mathrm{pp}}^{\mathrm{elastic}} = 10.35$\,mb from
  PYTHIA, summing elastic and single-diffractive contributions.
\item Boundary conditions used in the integration of
  Eq.~\eqref{eq:pbremXS}: by varying the upper bound on $p_{\perp}$ by
  $\pm 2$\,GeV, the total rate is changed by
  $_{-30\%}^{+15\%}$. Varying the lower and upper bounds
  $p_{\mathrm{min}}$ ($p_{\mathrm{max}}$) by $\pm 0.04$ ($\mp 0.04$),
  the total rate is changed by $_{-25\%}^{+40\%}$.
\end{itemize}
A total systematic uncertainty of $_{-40\%}^{+50\%}$ is assumed to
cover these sources.

For the QCD production, the theory systematic uncertainties concern
the parameterisation of the LO cross section, the choice of NNLO
k-factor and the impact from QCD scales and
PDFs. Figure~\ref{fig:qcdXS} shows the relative contributions from QCD
scales and PDF on PYTHIA's LO cross section. The choice of PDF set is
giving large variations in normalisation, but not affecting the
overall shape of the cross section versus mass. The PDF set chosen is
conservatively the one giving the lowest cross section. As discussed
in Section~\ref{sec:qcd}, in the end the uncertainty is dominated by
the NNLO k-factor of $1.7 \pm 0.7$. The total systematic uncertainty
is hence taken as $\pm 40\%$.

Experimental systematic uncertainties concern the measurement of the
luminosity, the modeling of the tracking efficiency and the
assumptions entering the 0-background estimate. They have been
neglected in this study, as they are expected to be small compared to
the theoretical uncertainties.

\subsection{Extraction of the limit}

Events are generated following a discrete grid in (\mDP,$\varepsilon$)
values, and passed through the full simulation of the SHiP detector
and reconstruction algorithms. The \DP mass is varied between the
electron-pair production threshold, and 10\,GeV, in 0.001 to 1\,GeV
steps. The kinetic mixing parameter $\varepsilon$ is varied between
$10^{-4}$ and $10^{-9}$ in varying-size steps in log($\varepsilon$).

To find the $\varepsilon$ values that allow to reach 2.3 expected
events, the expected rate is studied as a function of $\varepsilon$
for the discrete mass points, with a linear interpolation between
fully-simulated values. Between mass points, a linear interpolation is
also performed. The rate of events is driven by two aspects. For large
$\varepsilon$ values, larger cross sections are expected but the
detection efficiency decreases rapidly due to small lifetimes and
decays happening before the decay vessel. As $\varepsilon$ decreases,
the cross section decreases as $\varepsilon^2$ but the events have
more and more probability to reach the vessel and the rate increases,
up to a turning point where the decay vertex happens after the decay
vessel and/or the cross section becomes too small. Hence the 90\% CL
exclusion region is contained inside a lower and upper limits on
$\varepsilon^2$ for each mass point. The dependency of the excluded
region on the mass is driven by the kinematic properties of the \DP
and its decay products, affecting the detector acceptance and
selection efficiency.

\begin{figure}[h!]
  \centering
  \includegraphics[width=0.32\textwidth]{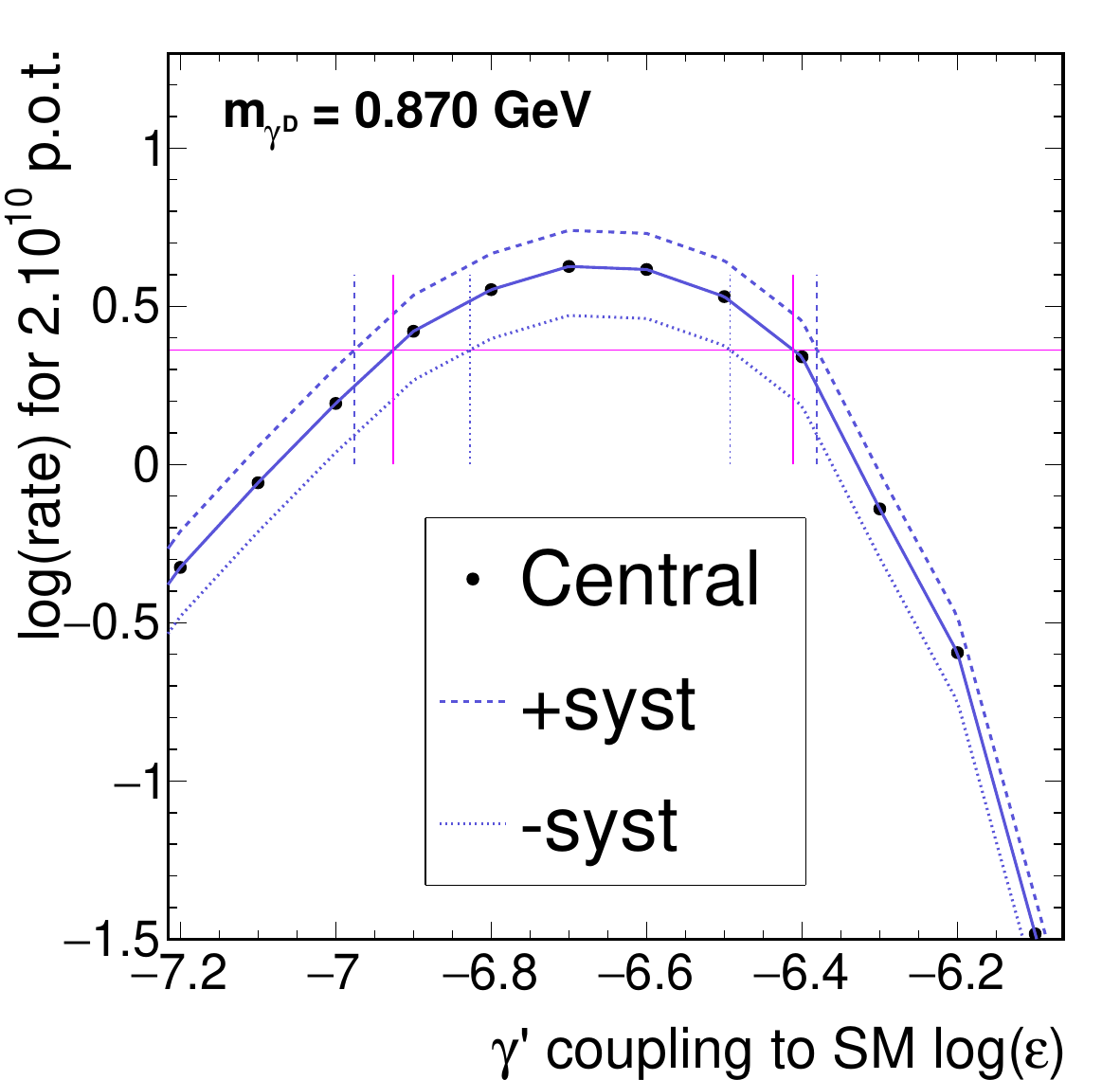}
  \hfill
  \includegraphics[width=0.32\textwidth]{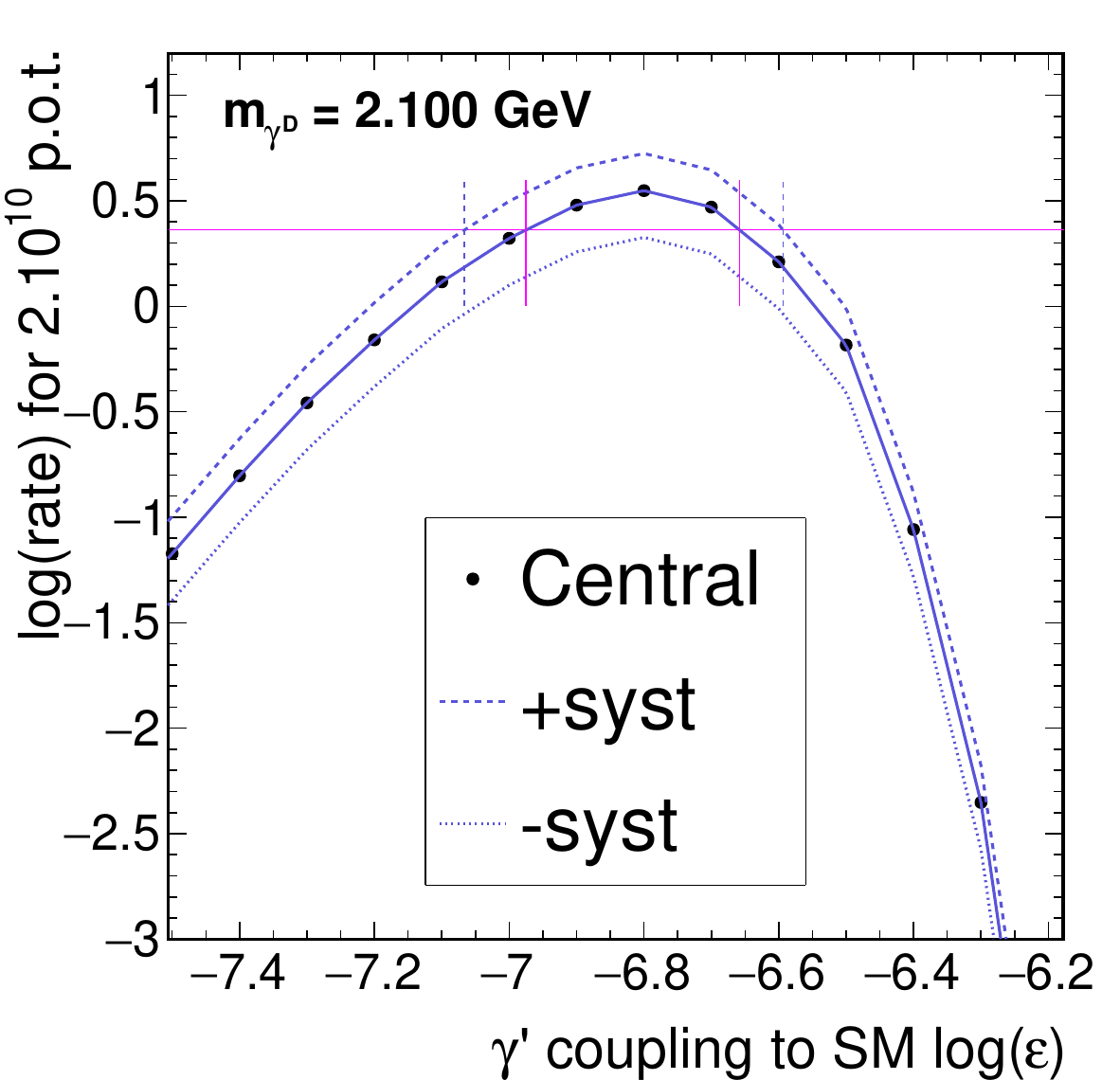}
  \hfill
  \includegraphics[width=0.32\textwidth]{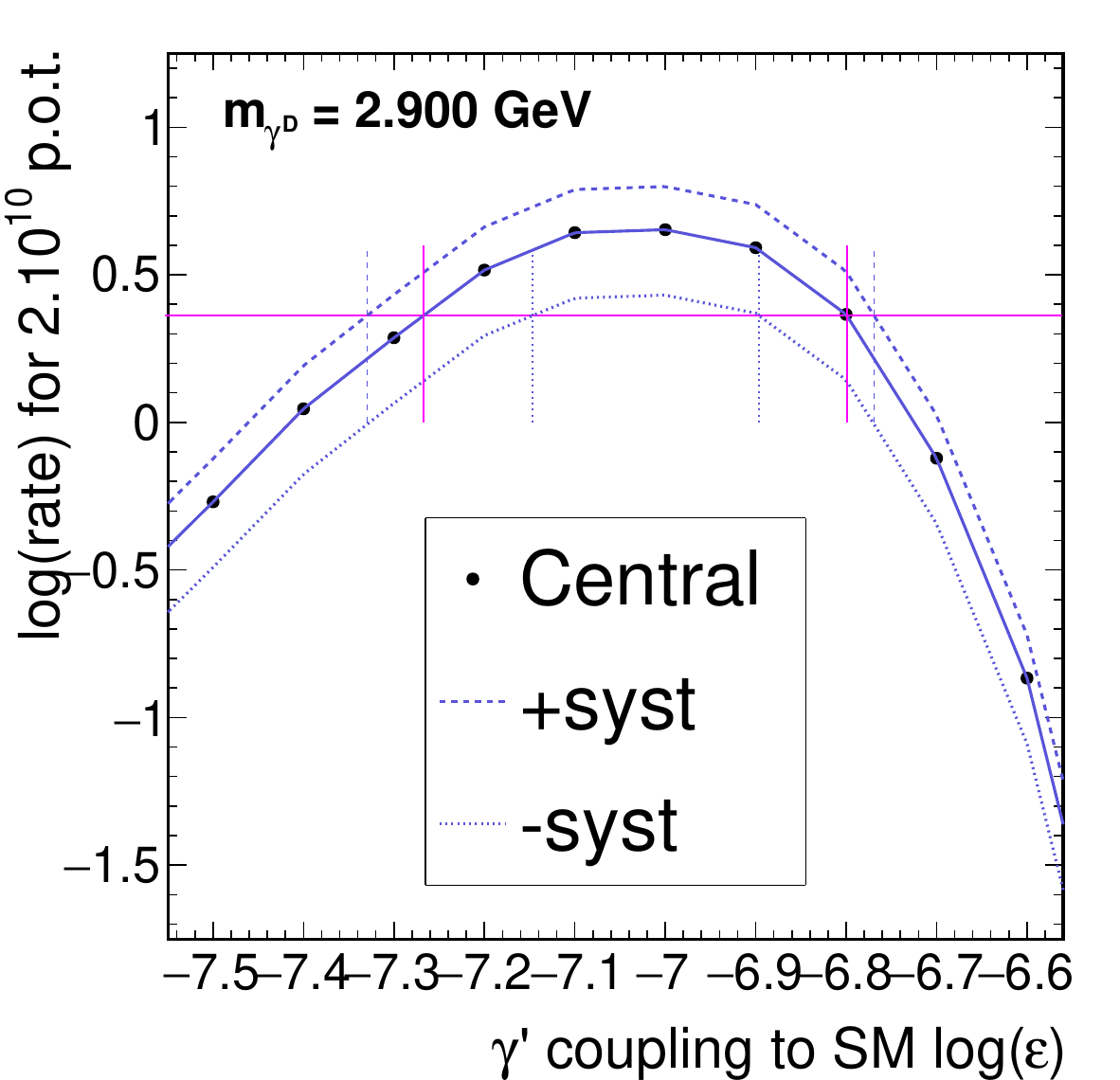}
\caption{Expected rate as a function of $\varepsilon$, for \mDP$=0.87$
  (left), 2.1 (middle) or 2.9 (right) GeV and meson, proton
  bremsstrahlung with dipole form factor or QCD production,
  respectively. The horizontal pink line shows the 2.3 events
  threshold used to set the limit. Vertical pink (dashed blue) lines
  show the result of the interpolation (varying the yields up and down
  according to the total systematic uncertainties detailed in
  Section~\ref{sec:systs}).}
\label{fig:ratevseps}
\end{figure}

As shown in Fig.~\ref{fig:ratevseps} for representative mass points,
for all processes, the upper bounds have little dependency on the
absolute normalisation of the rate (so in particular systematic
uncertainties on the cross sections and other quantities affecting the
overall rate), due to the very steep dependency of the rate as a
function of $\varepsilon$. The lower bounds are however more
sensitive.

\begin{figure}[h!]
  \centering
\includegraphics[width=1.\textwidth]{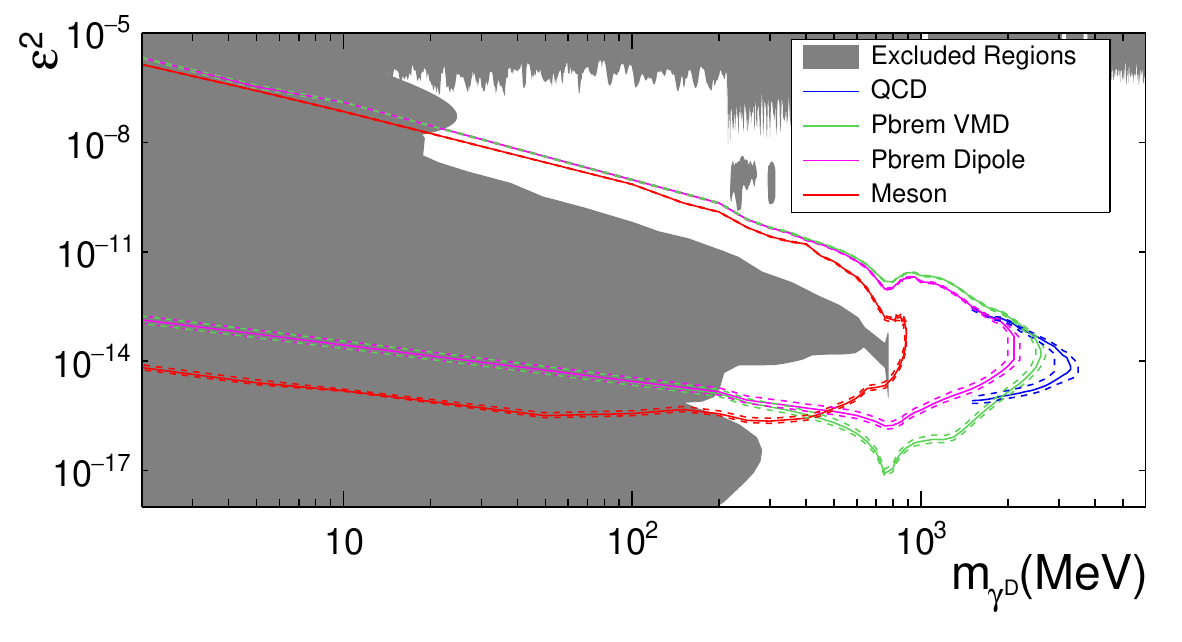}\\
\includegraphics[width=1.\textwidth]{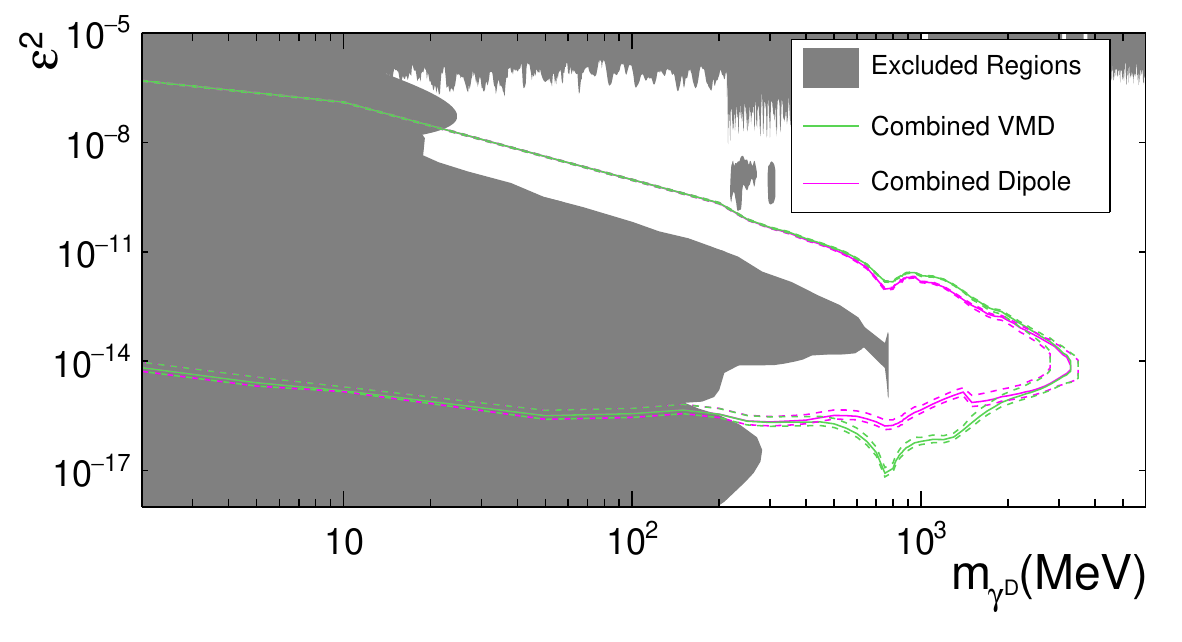}
\caption{Expected 90\% exclusion region as a function of the dark
  photon mass and of the kinetic mixing parameter $\varepsilon^2$, for
  the three production modes studied (top), and their combinations for
  the two proton bremsstrahlung scenarios (bottom). The dashed lines
  highlight the 1-$\sigma$ uncertainty band using the systematics
  described in Section~\ref{sec:systs}. The excluded region in grey is
  from Ref.~\cite{Beacham_2019}.}
\label{fig:sensitivity}
\end{figure}

The 90\% CL exclusion contour is shown in Fig.~\ref{fig:sensitivity}
for the three production modes studied, and their combinations, in the
(\mDP,$\varepsilon^2$) plane. The excluded region shown in grey is
from past experiments sensitive to this
process~\cite{Beacham_2019}. For $\varepsilon^2 > 10^{-6}$, in the
full mass range studied, the current sensitivity is coming from
searches for dilepton resonances
(e.g.~\cite{PhysRevLett.113.201801,2015178,PhysRevLett.112.221802,ARCHILLI2012251,BABUSCI2013111,2014459,2016356,PhysRevLett.120.061801}). These
results are complemented for low masses at lower $\varepsilon^2$
values by those from the reinterpretation of data from fixed-target
experiments
(e.g.~\cite{PhysRevD.38.3375,PhysRevLett.113.171802,BERGSMA1985458,PhysRevLett.59.755,PhysRevLett.67.2942}),
and by recent dedicated searches for long-lived \DP decaying to
leptons~\cite{Aaij:2019bvg,Banerjee:2019hmi}. The very-low coupling
exclusions are from cosmological constraints, in particular bounds
from Supernova 1987A
data~\cite{Fradette:2014sza,Kazanas:2014mca,Chang:2016ntp}.

The SHiP experiment is expected to have a unique sensitivity in the
mass region m$_{\gamma^{\mathrm{D}}}$ ranging between 0.8 and 3.3$^{+0.2}_{-0.5}$\,GeV,
and $\varepsilon^2$ ranging between $10^{-11}$ and $10^{-17}$.

\section{Conclusion}
\label{sec:concl}

The sensitivity of the SHiP detector has been investigated for the
simplest vector portal model, in which the only hidden-sector particle
connecting to SM particles is a dark photon. The model is fully
parameterised by only two parameters, the mass of the dark photon \mDP
and the kinetic mixing parameter $\varepsilon$. Three different
production mechanisms have been investigated, namely the production
via meson decays from non-diffractive proton-nucleon interactions, by
proton bremsstrahlung and by QCD parton-parton interaction. Different
sources of systematic uncertainties have been considered, dominated by
theory predictions on the cross section times branching ratios (meson
decays), two scenarios of nuclear form factor (proton bremsstrahlung)
and higher-order corrections (QCD scattering). Only the primary
proton-nucleon interaction is taken into account, secondaries from
hadronic interactions in cascade decays could lead to an improvement
in the sensitivity and will be the object of future work. The dark
photon is assumed to decay to pairs of leptons or quarks, and only
decay channels producing at least two charged particles coming from a
common vertex are used. With the selection applied, backgrounds are
neglected and 90\% CL exclusion contours are extracted and compared
with those from past experiments. The SHiP detector is expected to
have a unique sensitivity for m$_{\gamma^{\mathrm{D}}}$ ranging
between 0.8 and 3.3$^{+0.2}_{-0.5}$\,GeV, and $\varepsilon^2$ ranging
between $10^{-11}$ and $10^{-17}$.

\section*{Acknowledgments}

The SHiP Collaboration wishes to thank the Castaldo company (Naples,
Italy) for their contribution to the development studies of the decay
vessel. The SHiP Collaboration acknowledges support from the following
funding agencies: the TAEK of Turkey with grant number
2018TAEK(CERN)-A5.H6.F2-17; the National Research Foundation of Korea
with grant numbers 2018R1A2B2007757, 2018R1D1A3B07050649,
2018R1D1A1B07050701, 2017R1D1A1B03036042, 2017R1A6A3A01075752,
2016R1A2B4012302, and 2016R1A6A3A11930680; the Funda\c{c}\~{a}o para a
Ci\^{e}ncia e a Tecnologia of Portugal with grant number
CERN/FIS-PAR/0030/2017; the Russian Foundation for Basic Research
(RFBR), with grant number 17-02-00607.


\bibliographystyle{JHEP}

\bibliography{dp_paper}

\providecommand{\href}[2]{#2}\begingroup\raggedright\begin{thebibliography}{10}

\bibitem{LHC}
L.~Evans and P.~Bryant, \emph{{LHC Machine}},
  \href{https://doi.org/10.1088/1748-0221/3/08/S08001}{\emph{JINST} {\bfseries
  3} (2008) S08001}.

\bibitem{Apollinari:2015bam}
G.~Apollinari, I.~Béjar~Alonso, O.~Brüning, M.~Lamont and L.~Rossi,
  \emph{{High-Luminosity Large Hadron Collider (HL-LHC): Preliminary Design
  Report}},  Tech. Rep. CERN-2015-005, FERMILAB-DESIGN-2015-02, CERN, Geneva
  (2015).

\bibitem{Ahdida:2655435}
C.~Ahdida et~al., \emph{{A Beam Dump Facility (BDF) at CERN – The Concept and
  a First Radiological Assessment }},  Tech. Rep.
  \href{https://cds.cern.ch/record/2655435}{CERN-PBC-CONF-2019-001}, CERN,
  Geneva (Jan, 2019).

\bibitem{PhysRevD.80.095024}
B.~Batell, M.~Pospelov and A.~Ritz, \emph{Exploring portals to a hidden sector
  through fixed targets},
  \href{https://doi.org/10.1103/PhysRevD.80.095024}{\emph{Phys. Rev. D}
  {\bfseries 80} (2009) 095024}.

\bibitem{Alekhin_2016}
{\scshape SHiP} collaboration, \emph{A facility to search for hidden particles
  at the {CERN} {SPS}: the {SHiP} physics case},
  \href{https://doi.org/10.1088/0034-4885/79/12/124201}{\emph{Reports on
  Progress in Physics} {\bfseries 79} (2016) 124201}.

\bibitem{Patt:2006fw}
B.~Patt and F.~Wilczek, \emph{{Higgs-field portal into hidden sectors}},  Tech.
  Rep. \href{https://arxiv.org/abs/hep-ph/0605188}{MIT-CTP-3745} (5, 2006).

\bibitem{PhysRevD.75.037701}
D.~O'Connell, M.J.~Ramsey-Musolf and M.B.~Wise, \emph{Minimal extension of the
  standard model scalar sector},
  \href{https://doi.org/10.1103/PhysRevD.75.037701}{\emph{Phys. Rev. D}
  {\bfseries 75} (2007) 037701}.

\bibitem{Holdom:1985ag}
B.~Holdom, \emph{{Two U(1)'s and Epsilon Charge Shifts}},
  \href{https://doi.org/10.1016/0370-2693(86)91377-8}{\emph{Phys. Lett.}
  {\bfseries 166B} (1986) 196}.

\bibitem{Bauer:2018onh}
M.~Bauer, P.~Foldenauer and J.~Jaeckel, \emph{{Hunting All the Hidden
  Photons}}, \href{https://doi.org/10.1007/JHEP07(2018)094}{\emph{JHEP}
  {\bfseries 18} (2020) 094}
  [\href{https://arxiv.org/abs/1803.05466}{{\ttfamily 1803.05466}}].

\bibitem{Gorbunov_2007}
D.~Gorbunov and M.~Shaposhnikov, \emph{How to find neutral leptons of the
  $\upnu${MSM}?},
  \href{https://doi.org/10.1088/1126-6708/2007/10/015}{\emph{Journal of High
  Energy Physics} {\bfseries 2007} (2007) 015}.

\bibitem{Aaboud:2018fvk}
{\scshape ATLAS} collaboration, \emph{{Search for Higgs boson decays to
  beyond-the-Standard-Model light bosons in four-lepton events with the ATLAS
  detector at $\sqrt{s}=13$ TeV}},
  \href{https://doi.org/10.1007/JHEP06(2018)166}{\emph{JHEP} {\bfseries 06}
  (2018) 166} [\href{https://arxiv.org/abs/1802.03388}{{\ttfamily
  1802.03388}}].

\bibitem{Aaboud:2017buh}
{\scshape ATLAS} collaboration, \emph{{Search for new high-mass phenomena in
  the dilepton final state using 36 fb$^{-1}$ of proton-proton collision data
  at $ \sqrt{s}=13 $ TeV with the ATLAS detector}},
  \href{https://doi.org/10.1007/JHEP10(2017)182}{\emph{JHEP} {\bfseries 10}
  (2017) 182} [\href{https://arxiv.org/abs/1707.02424}{{\ttfamily
  1707.02424}}].

\bibitem{Sirunyan:2018mgs}
{\scshape CMS} collaboration, \emph{{A search for pair production of new light
  bosons decaying into muons in proton-proton collisions at 13 TeV}},
  \href{https://doi.org/10.1016/j.physletb.2019.07.013}{\emph{Phys. Lett. B}
  {\bfseries 796} (2019) 131}
  [\href{https://arxiv.org/abs/1812.00380}{{\ttfamily 1812.00380}}].

\bibitem{Khachatryan:2016zqb}
{\scshape CMS} collaboration, \emph{{Search for narrow resonances in dilepton
  mass spectra in proton-proton collisions at $\sqrt{s}$ = 13 TeV and
  combination with 8 TeV data}},
  \href{https://doi.org/10.1016/j.physletb.2017.02.010}{\emph{Phys. Lett.}
  {\bfseries B768} (2017) 57}
  [\href{https://arxiv.org/abs/1609.05391}{{\ttfamily 1609.05391}}].

\bibitem{Anelli:2007512}
{\scshape SHiP Collaboration} collaboration, \emph{{A Facility to Search for
  Hidden Particles (SHiP) at the CERN SPS}},  Tech. Rep.
  \href{https://cds.cern.ch/record/2007512}{CERN-SPSC-2015-016. SPSC-P-350.
  arXiv:1504.04956}, CERN, Geneva (Apr, 2015).

\bibitem{Bonivento:1606085}
W.~Bonivento et~al., \emph{{Proposal to Search for Heavy Neutral Leptons at the
  SPS}},  Tech. Rep.
  \href{http://cds.cern.ch/record/1606085}{CERN-SPSC-2013-024. SPSC-EOI-010},
  CERN, Geneva (Oct, 2013).

\bibitem{SHiP:2018xqw}
{\scshape SHiP} collaboration, \emph{{Sensitivity of the SHiP experiment to
  Heavy Neutral Leptons}},
  \href{https://doi.org/10.1007/JHEP04(2019)077}{\emph{JHEP} {\bfseries 04}
  (2019) 077} [\href{https://arxiv.org/abs/1811.00930}{{\ttfamily
  1811.00930}}].

\bibitem{SHiP:2020hyy}
{\scshape SHiP} collaboration, \emph{{Measurement of the muon flux for the SHiP
  experiment}},  Tech. Rep.
  \href{https://arxiv.org/abs/2001.04784}{CERN-EP-2020-006}, CERN, Geneva (1,
  2020).

\bibitem{Akmete:2017bpl}
{\scshape SHiP} collaboration, \emph{{The active muon shield in the SHiP
  experiment}},
  \href{https://doi.org/10.1088/1748-0221/12/05/P05011}{\emph{JINST} {\bfseries
  12} (2017) P05011} [\href{https://arxiv.org/abs/1703.03612}{{\ttfamily
  1703.03612}}].

\bibitem{Ahdida:2704147}
{\scshape SHiP} collaboration, \emph{{SHiP Experiment - Comprehensive Design
  Study report}},  Tech. Rep.
  \href{https://cds.cern.ch/record/2704147}{CERN-SPSC-2019-049. SPSC-SR-263},
  CERN, Geneva (Dec, 2019).

\bibitem{Miano2020}
A.~Miano, A.~Fiorillo, A.~Salzano, A.~Prota and R.~Jacobsson, \emph{The
  structural design of the decay volume for the search for hidden particles
  (ship) project},
  \href{https://doi.org/10.1007/s43452-020-00152-9}{\emph{Archiv. Civ. Mech.
  Eng} {\bfseries 21} (2020) 3}.

\bibitem{Bereziuk:2005286}
{\scshape SHiP} collaboration, \emph{{Initial design studies of the SHiP straw
  detector}},  Tech. Rep.
  \href{https://cds.cern.ch/record/2005286}{CERN-SHiP-NOTE-2015-001}, CERN,
  Geneva (Mar, 2015).

\bibitem{VanHerwijnen:2005715}
{\scshape SHiP} collaboration, \emph{{Simulation and pattern recognition for
  the SHiP Spectrometer Tracker}},  Tech. Rep.
  \href{https://cds.cern.ch/record/2005715}{CERN-SHiP-NOTE-2015-002}, CERN,
  Geneva (Mar, 2015).

\bibitem{Hosseini:2282039}
{\scshape SHiP} collaboration, \emph{{Particle Identification tools and
  performance in the SHiP Experiment}},  Tech. Rep.
  \href{https://cds.cern.ch/record/2282039}{CERN-SHiP-NOTE-2017-002}, CERN,
  Geneva (Jul, 2017).

\bibitem{Sjostrand:2014zea}
T.~Sjöstrand, S.~Ask, J.R.~Christiansen, R.~Corke, N.~Desai, P.~Ilten et~al.,
  \emph{{An Introduction to PYTHIA 8.2}},
  \href{https://doi.org/10.1016/j.cpc.2015.01.024}{\emph{Comput. Phys. Commun.}
  {\bfseries 191} (2015) 159}
  [\href{https://arxiv.org/abs/1410.3012}{{\ttfamily 1410.3012}}].

\bibitem{Genie}
C.~Andreopoulos et~al., \emph{{The GENIE Neutrino Monte Carlo Generator}},
  \href{https://doi.org/10.1016/j.nima.2009.12.009}{\emph{Nucl. Instrum. Meth.}
  {\bfseries A614} (2010) 87}
  [\href{https://arxiv.org/abs/0905.2517}{{\ttfamily 0905.2517}}].

\bibitem{pythia6}
T.~Sjöstrand, S.~Mrenna and P.Z.~Skands, \emph{Pythia 6.4 physics and manual},
  {\emph{Journal of High Energy Physics} {\bfseries 2006} (2006) 026}.

\bibitem{CERN-SHiP-NOTE-2015-009}
{\scshape SHiP} collaboration, \emph{{Heavy Flavour Cascade Production in a
  Beam Dump}},  Tech. Rep.
  \href{https://cds.cern.ch/record/2115534}{CERN-SHiP-NOTE-2015-009}, CERN,
  Geneva (Dec, 2015).

\bibitem{Geant4}
{\scshape GEANT4} collaboration, \emph{{GEANT4: A Simulation toolkit}},
  \href{https://doi.org/10.1016/S0168-9002(03)01368-8}{\emph{Nucl. Instrum.
  Meth.} {\bfseries A506} (2003) 250}.

\bibitem{FAIRROOT}
M.~Al-Turany, D.~Bertini, R.~Karabowicz, D.~Kresan, P.~Malzacher, T.~Stockmanns
  et~al., \emph{{The FairRoot framework}},
  \href{https://doi.org/10.1088/1742-9596/396/2/022001}{\emph{Journal of
  Physics: Conference Series} {\bfseries 396} (2012) 022001}.

\bibitem{Gorbunov:2014wqa}
D.~Gorbunov, A.~Makarov and I.~Timiryasov, \emph{{Decaying light particles in
  the SHiP experiment: Signal rate estimates for hidden photons}},
  \href{https://doi.org/10.1103/PhysRevD.91.035027}{\emph{Phys. Rev.}
  {\bfseries D91} (2015) 035027}
  [\href{https://arxiv.org/abs/1411.4007}{{\ttfamily 1411.4007}}].

\bibitem{Ball:2012cx}
R.D.~Ball et~al., \emph{{Parton distributions with LHC data}},
  \href{https://doi.org/10.1016/j.nuclphysb.2012.10.003}{\emph{Nucl. Phys.}
  {\bfseries B867} (2013) 244}
  [\href{https://arxiv.org/abs/1207.1303}{{\ttfamily 1207.1303}}].

\bibitem{Skands:2014pea}
P.~Skands, S.~Carrazza and J.~Rojo, \emph{{Tuning PYTHIA 8.1: the Monash 2013
  Tune}}, \href{https://doi.org/10.1140/epjc/s10052-014-3024-y}{\emph{Eur.
  Phys. J. C} {\bfseries 74} (2014) 3024}
  [\href{https://arxiv.org/abs/1404.5630}{{\ttfamily 1404.5630}}].

\bibitem{Batell:2009di}
B.~Batell, M.~Pospelov and A.~Ritz, \emph{{Exploring Portals to a Hidden Sector
  Through Fixed Targets}},
  \href{https://doi.org/10.1103/PhysRevD.80.095024}{\emph{Phys. Rev.}
  {\bfseries D80} (2009) 095024}
  [\href{https://arxiv.org/abs/0906.5614}{{\ttfamily 0906.5614}}].

\bibitem{Berlin:2018pwi}
A.~Berlin, S.~Gori, P.~Schuster and N.~Toro, \emph{{Dark Sectors at the
  Fermilab SeaQuest Experiment}},
  \href{https://doi.org/10.1103/PhysRevD.98.035011}{\emph{Phys. Rev.}
  {\bfseries D98} (2018) 035011}
  [\href{https://arxiv.org/abs/1804.00661}{{\ttfamily 1804.00661}}].

\bibitem{Blumlein:2013cua}
J.~Blümlein and J.~Brunner, \emph{{New Exclusion Limits on Dark Gauge Forces
  from Proton Bremsstrahlung in Beam-Dump Data}},
  \href{https://doi.org/10.1016/j.physletb.2014.02.029}{\emph{Phys. Lett.}
  {\bfseries B731} (2014) 320}
  [\href{https://arxiv.org/abs/1311.3870}{{\ttfamily 1311.3870}}].

\bibitem{Patrignani:2016xqp}
{\scshape Particle Data Group} collaboration, \emph{{Review of Particle
  Physics}}, \href{https://doi.org/10.1088/1674-1137/40/10/100001}{\emph{Chin.
  Phys.} {\bfseries C40} (2016) 100001}.

\bibitem{RevModPhys.35.335}
L.N.~Hand, D.G.~Miller and R.~Wilson, \emph{Electric and magnetic form factors
  of the nucleon}, \href{https://doi.org/10.1103/RevModPhys.35.335}{\emph{Rev.
  Mod. Phys.} {\bfseries 35} (1963) 335}.

\bibitem{deNiverville:2016rqh}
P.~deNiverville, C.-Y.~Chen, M.~Pospelov and A.~Ritz, \emph{{Light dark matter
  in neutrino beams: production modelling and scattering signatures at
  MiniBooNE, T2K and SHiP}},
  \href{https://doi.org/10.1103/PhysRevD.95.035006}{\emph{Phys. Rev.}
  {\bfseries D95} (2017) 035006}
  [\href{https://arxiv.org/abs/1609.01770}{{\ttfamily 1609.01770}}].

\bibitem{Faessler:2009tn}
A.~Faessler, M.I.~Krivoruchenko and B.V.~Martemyanov, \emph{{Once more on
  electromagnetic form factors of nucleons in extended vector meson dominance
  model}}, \href{https://doi.org/10.1103/PhysRevC.82.038201}{\emph{Phys. Rev.}
  {\bfseries C82} (2010) 038201}
  [\href{https://arxiv.org/abs/0910.5589}{{\ttfamily 0910.5589}}].

\bibitem{Nakamura_2010}
{\scshape Particle Data Group} collaboration, \emph{{Review of Particle
  Physics}},
  \href{https://doi.org/10.1088/0954-3899/37/7a/075021}{\emph{Journal of
  Physics G: Nuclear and Particle Physics} {\bfseries 37} (2010) 075021}.

\bibitem{Kim_Tsai_1973}
K.~Kim and Y.~Tsai, \emph{{Improved Weizs\"{a}cker-Williams method and its
  application to lepton and W boson pair production}}, {\emph{Phys. Rev.}
  {\bfseries D8} (1973) 3109}.

\bibitem{KIM1972665}
K.~Kim and Y.~Tsai, \emph{{An improved Weizs\"{a}cker-Williams method and
  photoproduction of lepton pairs}},
  \href{https://doi.org/10.1016/0370-2693(72)90622-3}{\emph{Physics Letters B}
  {\bfseries 40} (1972) 665}.

\bibitem{Carloni:2011kk}
L.~Carloni, J.~Rathsman and T.~Sjöstrand, \emph{{Discerning Secluded Sector
  gauge structures}},
  \href{https://doi.org/10.1007/JHEP04(2011)091}{\emph{JHEP} {\bfseries 04}
  (2011) 091} [\href{https://arxiv.org/abs/1102.3795}{{\ttfamily 1102.3795}}].

\bibitem{Ciobanu:2005pv}
{\scshape CDF} collaboration, \emph{{Z' generation with PYTHIA}},  Tech. Rep.
  FERMILAB-FN-0773-E, FERMILAB, Batavia (2005),
  \href{https://doi.org/10.2172/15020136}{DOI}.

\bibitem{Buckley:2014ana}
A.~Buckley, J.~Ferrando, S.~Lloyd, K.~Nordstr\"om, B.~Page, M.~R\"ufenacht
  et~al., \emph{{LHAPDF6: parton density access in the LHC precision era}},
  \href{https://doi.org/10.1140/epjc/s10052-015-3318-8}{\emph{Eur. Phys. J. C}
  {\bfseries 75} (2015) 132} [\href{https://arxiv.org/abs/1412.7420}{{\ttfamily
  1412.7420}}].

\bibitem{Eskola:2016oht}
K.J.~Eskola, P.~Paakkinen, H.~Paukkunen and C.A.~Salgado, \emph{{EPPS16:
  Nuclear parton distributions with LHC data}},
  \href{https://doi.org/10.1140/epjc/s10052-017-4725-9}{\emph{Eur. Phys. J. C}
  {\bfseries 77} (2017) 163}
  [\href{https://arxiv.org/abs/1612.05741}{{\ttfamily 1612.05741}}].

\bibitem{Alwall:2014hca}
J.~Alwall, R.~Frederix, S.~Frixione, V.~Hirschi, F.~Maltoni, O.~Mattelaer
  et~al., \emph{{The automated computation of tree-level and next-to-leading
  order differential cross sections, and their matching to parton shower
  simulations}}, \href{https://doi.org/10.1007/JHEP07(2014)079}{\emph{JHEP}
  {\bfseries 07} (2014) 079} [\href{https://arxiv.org/abs/1405.0301}{{\ttfamily
  1405.0301}}].

\bibitem{Grazzini:2017mhc}
M.~Grazzini, S.~Kallweit and M.~Wiesemann, \emph{{Fully differential NNLO
  computations with MATRIX}},
  \href{https://doi.org/10.1140/epjc/s10052-018-5771-7}{\emph{Eur. Phys. J. C}
  {\bfseries 78} (2018) 537}
  [\href{https://arxiv.org/abs/1711.06631}{{\ttfamily 1711.06631}}].

\bibitem{Catani:2009sm}
S.~Catani, L.~Cieri, G.~Ferrera, D.~de~Florian and M.~Grazzini, \emph{{Vector
  boson production at hadron colliders: a fully exclusive QCD calculation at
  NNLO}}, \href{https://doi.org/10.1103/PhysRevLett.103.082001}{\emph{Phys.
  Rev. Lett.} {\bfseries 103} (2009) 082001}
  [\href{https://arxiv.org/abs/0903.2120}{{\ttfamily 0903.2120}}].

\bibitem{Cascioli:2011va}
F.~Cascioli, P.~Maierh\"ofer and S.~Pozzorini, \emph{{Scattering Amplitudes
  with Open Loops}},
  \href{https://doi.org/10.1103/PhysRevLett.108.111601}{\emph{Phys. Rev. Lett.}
  {\bfseries 108} (2012) 111601}
  [\href{https://arxiv.org/abs/1111.5206}{{\ttfamily 1111.5206}}].

\bibitem{Denner:2016kdg}
A.~Denner, S.~Dittmaier and L.~Hofer, \emph{{Collier: a fortran-based Complex
  One-Loop LIbrary in Extended Regularizations}},
  \href{https://doi.org/10.1016/j.cpc.2016.10.013}{\emph{Comput. Phys. Commun.}
  {\bfseries 212} (2017) 220}
  [\href{https://arxiv.org/abs/1604.06792}{{\ttfamily 1604.06792}}].

\bibitem{Catani:2012qa}
S.~Catani, L.~Cieri, D.~de~Florian, G.~Ferrera and M.~Grazzini, \emph{{Vector
  boson production at hadron colliders: hard-collinear coefficients at the
  NNLO}}, \href{https://doi.org/10.1140/epjc/s10052-012-2195-7}{\emph{Eur.
  Phys. J.} {\bfseries C72} (2012) 2195}
  [\href{https://arxiv.org/abs/1209.0158}{{\ttfamily 1209.0158}}].

\bibitem{Catani:2007vq}
S.~Catani and M.~Grazzini, \emph{{An NNLO subtraction formalism in hadron
  collisions and its application to Higgs boson production at the LHC}},
  \href{https://doi.org/10.1103/PhysRevLett.98.222002}{\emph{Phys. Rev. Lett.}
  {\bfseries 98} (2007) 222002}
  [\href{https://arxiv.org/abs/hep-ph/0703012}{{\ttfamily hep-ph/0703012}}].

\bibitem{Martin:2009iq}
A.D.~Martin, W.J.~Stirling, R.S.~Thorne and G.~Watt, \emph{{Parton
  distributions for the LHC}},
  \href{https://doi.org/10.1140/epjc/s10052-009-1072-5}{\emph{Eur. Phys. J. C}
  {\bfseries 63} (2009) 189} [\href{https://arxiv.org/abs/0901.0002}{{\ttfamily
  0901.0002}}].

\bibitem{Shimizu:2005fp}
H.~Shimizu, G.F.~Sterman, W.~Vogelsang and H.~Yokoya, \emph{{Dilepton
  production near partonic threshold in transversely polarized
  proton-antiproton collisions}},
  \href{https://doi.org/10.1103/PhysRevD.71.114007}{\emph{Phys. Rev. D}
  {\bfseries 71} (2005) 114007}
  [\href{https://arxiv.org/abs/hep-ph/0503270}{{\ttfamily hep-ph/0503270}}].

\bibitem{Ball:2017nwa}
{\scshape NNPDF} collaboration, \emph{{Parton distributions from high-precision
  collider data}},
  \href{https://doi.org/10.1140/epjc/s10052-017-5199-5}{\emph{Eur. Phys. J. C}
  {\bfseries 77} (2017) 663}
  [\href{https://arxiv.org/abs/1706.00428}{{\ttfamily 1706.00428}}].

\bibitem{Lai:2009ne}
H.-L.~Lai, J.~Huston, S.~Mrenna, P.~Nadolsky, D.~Stump, W.-K.~Tung et~al.,
  \emph{{Parton Distributions for Event Generators}},
  \href{https://doi.org/10.1007/JHEP04(2010)035}{\emph{JHEP} {\bfseries 04}
  (2010) 035} [\href{https://arxiv.org/abs/0910.4183}{{\ttfamily 0910.4183}}].

\bibitem{Pumplin:2002vw}
J.~Pumplin, D.R.~Stump, J.~Huston, H.L.~Lai, P.M.~Nadolsky and W.K.~Tung,
  \emph{{New generation of parton distributions with uncertainties from global
  QCD analysis}},
  \href{https://doi.org/10.1088/1126-6708/2002/07/012}{\emph{JHEP} {\bfseries
  07} (2002) 012} [\href{https://arxiv.org/abs/hep-ph/0201195}{{\ttfamily
  hep-ph/0201195}}].

\bibitem{Alekhin:2017kpj}
S.~Alekhin, J.~Bl\"umlein, S.~Moch and R.~Placakyte, \emph{{Parton distribution
  functions, $\alpha_s$, and heavy-quark masses for LHC Run II}},
  \href{https://doi.org/10.1103/PhysRevD.96.014011}{\emph{Phys. Rev. D}
  {\bfseries 96} (2017) 014011}
  [\href{https://arxiv.org/abs/1701.05838}{{\ttfamily 1701.05838}}].

\bibitem{Bjorken:2009mm}
J.D.~Bjorken, R.~Essig, P.~Schuster and N.~Toro, \emph{{New Fixed-Target
  Experiments to Search for Dark Gauge Forces}},
  \href{https://doi.org/10.1103/PhysRevD.80.075018}{\emph{Phys. Rev.}
  {\bfseries D80} (2009) 075018}
  [\href{https://arxiv.org/abs/0906.0580}{{\ttfamily 0906.0580}}].

\bibitem{Agashe:2014kda}
{\scshape Particle Data Group} collaboration, \emph{{Review of Particle
  Physics}}, \href{https://doi.org/10.1088/1674-1137/38/9/090001}{\emph{Chin.
  Phys.} {\bfseries C38} (2014) 090001}.

\bibitem{Liu:2017ryd}
M.X.~Liu, \emph{{Prospects of direct search for dark photon and dark Higgs in
  SeaQuest/E1067 experiment at the Fermilab main injector}},
  \href{https://doi.org/10.1142/S0217732317300087}{\emph{Mod. Phys. Lett.}
  {\bfseries A32} (2017) 1730008}.

\bibitem{Dobrich:2019dxc}
B.~Döbrich, J.~Jaeckel and T.~Spadaro, \emph{{Light in the beam dump.
  Axion-Like Particle production from decay photons in proton beam-dumps}},
  \href{https://doi.org/10.1007/JHEP05(2019)213}{\emph{JHEP} {\bfseries 05}
  (2019) 213} [\href{https://arxiv.org/abs/1904.02091}{{\ttfamily
  1904.02091}}].

\bibitem{Acharya:2017tlv}
{\scshape ALICE} collaboration, \emph{{$\pi ^{0}$ and $\eta $ meson production
  in proton-proton collisions at $\sqrt{s}=8$ TeV}},
  \href{https://doi.org/10.1140/epjc/s10052-018-5612-8}{\emph{Eur. Phys. J. C}
  {\bfseries 78} (2018) 263}
  [\href{https://arxiv.org/abs/1708.08745}{{\ttfamily 1708.08745}}].

\bibitem{ALICE-PUBLIC-2018-004}
{\scshape ALICE} collaboration, \emph{{Production of $\omega(782)$ in pp
  collisions at $\sqrt{s}$ = 7 TeV}},  Tech. Rep.
  \href{http://cds.cern.ch/record/2316785}{ALICE-PUBLIC-2018-004}, CERN, Geneva
  (May, 2018).

\bibitem{Adare:2011ht}
{\scshape PHENIX} collaboration, \emph{{Production of $\omega$ mesons in $p+p$,
  d+Au, Cu+Cu, and Au+Au collisions at $\sqrt{s_NN}=200$ GeV}},
  \href{https://doi.org/10.1103/PhysRevC.84.044902}{\emph{Phys. Rev. C}
  {\bfseries 84} (2011) 044902}
  [\href{https://arxiv.org/abs/1105.3467}{{\ttfamily 1105.3467}}].

\bibitem{shipLDM}
{\scshape SHiP} collaboration, \emph{{Sensitivity of the SHiP experiment to
  light dark matter}},  Tech. Rep. \href{https://arxiv.org/abs/2010.11057}{},
  CERN (2020).

\bibitem{Beacham_2019}
J.~Beacham, C.~Burrage, D.~Curtin, A.~De~Roeck, J.~Evans, J.L.~Feng et~al.,
  \emph{Physics beyond colliders at cern: beyond the standard model working
  group report}, \href{https://doi.org/10.1088/1361-6471/ab4cd2}{\emph{Journal
  of Physics G: Nuclear and Particle Physics} {\bfseries 47} (2019) 010501}.

\bibitem{PhysRevLett.113.201801}
{\scshape BaBar} collaboration, \emph{Search for a dark photon in
  ${e}^{+}{e}^{\ensuremath{-}}$ collisions at babar},
  \href{https://doi.org/10.1103/PhysRevLett.113.201801}{\emph{Phys. Rev. Lett.}
  {\bfseries 113} (2014) 201801}.

\bibitem{2015178}
{\scshape NA48/2} collaboration, \emph{Search for the dark photon in $\pi^{0}$
  decays},
  \href{https://doi.org/https://doi.org/10.1016/j.physletb.2015.04.068}{\emph{Physics
  Letters B} {\bfseries 746} (2015) 178}.

\bibitem{PhysRevLett.112.221802}
{\scshape A1} collaboration, \emph{Search at the mainz microtron for light
  massive gauge bosons relevant for the muon $g\ensuremath{-}2$ anomaly},
  \href{https://doi.org/10.1103/PhysRevLett.112.221802}{\emph{Phys. Rev. Lett.}
  {\bfseries 112} (2014) 221802}.

\bibitem{ARCHILLI2012251}
{\scshape KLOE-2} collaboration, \emph{Search for a vector gauge boson in
  $\phi$ meson decays with the kloe detector},
  \href{https://doi.org/https://doi.org/10.1016/j.physletb.2011.11.033}{\emph{Physics
  Letters B} {\bfseries 706} (2012) 251}.

\bibitem{BABUSCI2013111}
{\scshape KLOE-2} collaboration, \emph{Limit on the production of a light
  vector gauge boson in $\phi$ meson decays with the kloe detector},
  \href{https://doi.org/https://doi.org/10.1016/j.physletb.2013.01.067}{\emph{Physics
  Letters B} {\bfseries 720} (2013) 111}.

\bibitem{2014459}
{\scshape KLOE-2} collaboration, \emph{Search for light vector boson production
  in ${e}^{+}{e}^{\ensuremath{-}}\rightarrow\mu^+\mu^{\ensuremath{-}}\gamma$
  interactions with the kloe experiment},
  \href{https://doi.org/https://doi.org/10.1016/j.physletb.2014.08.005}{\emph{Physics
  Letters B} {\bfseries 736} (2014) 459}.

\bibitem{2016356}
{\scshape KLOE-2} collaboration, \emph{Limit on the production of a new vector
  boson in ${e}^{+}{e}^{\ensuremath{-}}\rightarrow\upsilon\gamma$,
  $\upsilon\rightarrow\mu^+\mu^{\ensuremath{-}}$ with the kloe experiment},
  \href{https://doi.org/https://doi.org/10.1016/j.physletb.2016.04.019}{\emph{Physics
  Letters B} {\bfseries 757} (2016) 356}.

\bibitem{PhysRevLett.120.061801}
{\scshape LHCb} collaboration, \emph{Search for dark photons produced in 13 tev
  $pp$ collisions},
  \href{https://doi.org/10.1103/PhysRevLett.120.061801}{\emph{Phys. Rev. Lett.}
  {\bfseries 120} (2018) 061801}.

\bibitem{PhysRevD.38.3375}
J.D.~Bjorken, S.~Ecklund, W.R.~Nelson, A.~Abashian, C.~Church, B.~Lu et~al.,
  \emph{Search for neutral metastable penetrating particles produced in the
  slac beam dump}, \href{https://doi.org/10.1103/PhysRevD.38.3375}{\emph{Phys.
  Rev. D} {\bfseries 38} (1988) 3375}.

\bibitem{PhysRevLett.113.171802}
B.~Batell, R.~Essig and Z.~Surujon, \emph{Strong constraints on sub-gev dark
  sectors from slac beam dump e137},
  \href{https://doi.org/10.1103/PhysRevLett.113.171802}{\emph{Phys. Rev. Lett.}
  {\bfseries 113} (2014) 171802}.

\bibitem{BERGSMA1985458}
{\scshape CHARM} collaboration, \emph{Search for axion-like particle production
  in 400 gev proton-copper interactions},
  \href{https://doi.org/https://doi.org/10.1016/0370-2693(85)90400-9}{\emph{Physics
  Letters B} {\bfseries 157} (1985) 458}.

\bibitem{PhysRevLett.59.755}
E.M.~Riordan et~al., \emph{Search for short-lived axions in an
  electron-beam-dump experiment},
  \href{https://doi.org/10.1103/PhysRevLett.59.755}{\emph{Phys. Rev. Lett.}
  {\bfseries 59} (1987) 755}.

\bibitem{PhysRevLett.67.2942}
A.~Bross, M.~Crisler, S.~Pordes, J.~Volk, S.~Errede and J.~Wrbanek,
  \emph{Search for short-lived particles produced in an electron beam dump},
  \href{https://doi.org/10.1103/PhysRevLett.67.2942}{\emph{Phys. Rev. Lett.}
  {\bfseries 67} (1991) 2942}.

\bibitem{Aaij:2019bvg}
{\scshape LHCb} collaboration, \emph{{Search for $A'\to\mu^+\mu^-$ Decays}},
  \href{https://doi.org/10.1103/PhysRevLett.124.041801}{\emph{Phys. Rev. Lett.}
  {\bfseries 124} (2020) 041801}
  [\href{https://arxiv.org/abs/1910.06926}{{\ttfamily 1910.06926}}].

\bibitem{Banerjee:2019hmi}
{\scshape NA64} collaboration, \emph{{Improved limits on a hypothetical X(16.7)
  boson and a dark photon decaying into $e^+e^-$ pairs}},
  \href{https://doi.org/10.1103/PhysRevD.101.071101}{\emph{Phys. Rev. D}
  {\bfseries 101} (2020) 071101}
  [\href{https://arxiv.org/abs/1912.11389}{{\ttfamily 1912.11389}}].

\bibitem{Fradette:2014sza}
A.~Fradette, M.~Pospelov, J.~Pradler and A.~Ritz, \emph{{Cosmological
  Constraints on Very Dark Photons}},
  \href{https://doi.org/10.1103/PhysRevD.90.035022}{\emph{Phys. Rev. D}
  {\bfseries 90} (2014) 035022}
  [\href{https://arxiv.org/abs/1407.0993}{{\ttfamily 1407.0993}}].

\bibitem{Kazanas:2014mca}
D.~Kazanas, R.N.~Mohapatra, S.~Nussinov, V.L.~Teplitz and Y.~Zhang,
  \emph{{Supernova Bounds on the Dark Photon Using its Electromagnetic Decay}},
  \href{https://doi.org/10.1016/j.nuclphysb.2014.11.009}{\emph{Nucl. Phys. B}
  {\bfseries 890} (2014) 17} [\href{https://arxiv.org/abs/1410.0221}{{\ttfamily
  1410.0221}}].

\bibitem{Chang:2016ntp}
J.H.~Chang, R.~Essig and S.D.~McDermott, \emph{{Revisiting Supernova 1987A
  Constraints on Dark Photons}},
  \href{https://doi.org/10.1007/JHEP01(2017)107}{\emph{JHEP} {\bfseries 01}
  (2017) 107} [\href{https://arxiv.org/abs/1611.03864}{{\ttfamily
  1611.03864}}].

\end{thebibliography}\endgroup



%
\flushleft{\Large\bf The SHiP Collaboration}
\vspace*{1mm}
\begin{flushleft}
C.~Ahdida$^{44}$,
A.~Akmete$^{48}$,
R.~Albanese$^{14,d,h}$,
A.~Alexandrov$^{14,32,34,d}$,
A.~Anokhina$^{39,m}$,
S.~Aoki$^{18}$,
G.~Arduini$^{44}$,
E.~Atkin$^{38}$,
N.~Azorskiy$^{29}$,
J.J.~Back$^{54}$
% ...

\vspace*{1cm}

{\footnotesize \it

% ...
$ ^{14}$Sezione INFN di Napoli, Napoli, Italy\\
% ...
$ ^{18}$Kobe University, Kobe, Japan\\
% ...
$ ^{29}$Joint Institute for Nuclear Research (JINR), Dubna, Russia\\
% ...
$ ^{32}$P.N.~Lebedev Physical Institute (LPI RAS), Moscow, Russia\\
% ...
$ ^{34}$National University of Science and Technology ``MISiS``, Moscow, Russia\\
% ...
$ ^{38}$National Research Nuclear University (MEPhI), Moscow, Russia\\
$ ^{39}$Skobeltsyn Institute of Nuclear Physics of Moscow State University (SINP MSU), Moscow, Russia\\
% ...
$ ^{44}$European Organization for Nuclear Research (CERN), Geneva, Switzerland\\
% ...
$ ^{48}$Middle East Technical University (METU), Ankara, Turkey\\
% ...
$ ^{54}$University of Warwick, Warwick, United Kingdom\\
% ...
$ ^{d}$Universit\`{a} di Napoli ``Federico II``, Napoli, Italy\\
% ...
$ ^{h}$Consorzio CREATE, Napoli, Italy\\
% ...
$ ^{m}$Also at Faculty of Physics M.V. Lomonosov Moscow State University, Russia\\
}
\end{flushleft}


\end{document}